\documentclass[preprints,article,accept,moreauthors,pdftex]{mdpi}
\firstpage{1} 
\pubvolume{1}
\issuenum{1}
\articlenumber{0}
\pubyear{2022}
\copyrightyear{2022}
\datereceived{} 
\dateaccepted{} 
\datepublished{} 
\hreflink{https://doi.org/} 

\Title{Modeling Reconstructed Images of Jets Launched by SANE Super-Eddington Accretion Flows Around SMBHs with the ngEHT}

\TitleCitation{Modeling Reconstructed Images of Jets Launched by SANE Super-Eddington Accretion Flows Around SMBHs with the ngEHT}

\Author{Brandon Curd$^{1,2,*}$, Razieh  Emami $^{2}$, Freek Roelofs$^{1,2}$ and Richard Anantua $^{3}$}

\AuthorNames{Brandon Curd, Razieh Emami, Freek Roelofs and Richard Anantua}

\AuthorCitation{Curd, B. et al.,
}

\address{%
$^{1}$ \quad Black Hole Initiative at Harvard University, 20 Garden Street, Cambridge, MA 02138, USA\\
$^{2}$ \quad Center for Astrophysics $\vert$ Harvard \& Smithsonian, 60 Garden Street, Cambridge, MA 02138, USA \\
$^{3}$ \quad Department of Physics $\&$ Astronomy, The University of Texas at San Antonio, One UTSA Circle, San Antonio, TX 78249, USA\\}
\corres{Correspondence: brandon.curd@cfa.harvard.edu
}

\abstract{ Tidal disruption events (TDEs) around super massive black holes (SMBHs) are a potential laboratory to study super-Eddington accretion disks and sometimes result in powerful jets or outflows which may shine in the radio and sub millimeter bands. In this work, we model the thermal synchrotron emission of jets from general relativistic radiation magneto-hydrodynamics (GRRMHD) simulations of a BH accretion disk/jet system which assumes the TDE resulted in a magnetized accretion disk around a BH accreting at $\sim 12-25$ times the Eddington accretion rate. Through synthetic observations with the Next Generation Event Horizon Telescope (ngEHT) and an image reconstruction analysis, we demonstrate that TDE jets may provide compelling targets, within the context of the models explored in this work. In particular, we find that jets launched by a SANE super-Eddington disk around a spin $a_*=0.9$ reach the ngEHT detection threshold at large distances (up to 100 Mpc in this work). A two-temperature plasma in the jet or weaker jets, such as a spin $a_*=0$ model, requires a much closer distance as we demonstrate detection at 10 Mpc for limiting cases of $a_*=0,\,\mathcal{R}=1$ or $a_*=0.9,\, \mathcal{R}=20$. We also demonstrate that TDE jets may appear as superluminal sources if the BH is rapidly rotating and the jet is viewed nearly face on.
}


\begin{document}

\section{Introduction}
 
Tidal disruptions of stars by super massive black holes (SMBHs), or tidal disruption events (TDEs), have recently become a regularly observed transient phenomenon. Stars which enter the tidal radius 
\begin{equation}
 R_t \simeq 7\times10^{12} \left(\dfrac{M_{\rm{BH}}}{10^6 M_\odot}\right)^{1/3} \left(\dfrac{M_*}{M_\odot}\right)^{-1/3}\left(\dfrac{R_*}{R_\odot}\right) \rm{[cm]}
\end{equation}
of the central SMBH in their host galaxy will be disrupted \cite{1975Natur.254..295H,1988Natur.333..523R}, either partially or fully depending on the orbit and equation of state of the star \cite{2013ApJ...767...25G,2017A&A...600A.124M}. The bound stream of gas returns towards the BH delivering mass at the fallback rate ($\dot{M}_{\rm{fb}}$). Apsidal precession of the returning stream leads to self-intersection with material that has yet to pass through pericenter and leads to dissipation and disk formation. Dissipation due to self-intersection may also be a source of early emission in a TDE \cite{Steinberg20221}. After the initial rise to peak, the fallback rate follows a power law behaviour which can be approximated as
\begin{equation}
    \dot{M}_{\rm{fb}}=\dot{M}_{\rm{fb,peak}}\left(\dfrac{t}{t_{\rm{fb}}}\right)^{-5/3}.
\end{equation}
Here $\dot{M}_{\rm{fb,peak}}$ is the peak mass fallback rate
\begin{equation} \label{eq:mdotfbpeak}
  \dfrac{\dot{M}_{\rm{fb, peak}}}{\dot{M}_{\rm{Edd}}} \approx  133 \left(\dfrac{M_{\rm{BH}}}{10^6 M_\odot}\right)^{-3/2} \left(\dfrac{M_*}{M_\odot}\right)^{2}\left(\dfrac{R_*}{R_\odot}\right)^{-3/2}
\end{equation}
making the ``frozen in'' approximation as in \citet{2013MNRAS.435.1809S}, and
\begin{equation} \label{eq:tfb}
t_{\rm{fb}} = 3.5\times10^{6} \, {\rm{s}} \, \left(\dfrac{M_{\rm{BH}}}{10^6 M_\odot}\right)^{1/2} \left(\dfrac{M_*}{M_\odot}\right)^{-1}\left(\dfrac{R_*}{R_\odot}\right)^{3/2},
\end{equation}
is the fallback time, which is the orbital time of the most bound part of the stream. Of note is the fact that the mass fallback rate can greatly exceed the Eddington mass accretion rate $\dot{M}_{\rm{Edd}}$. The exact power law behaviour varies with the properties and orbit of the star \cite{2019ApJ...882L..26G}. 

TDEs are typically seen as optical/X-ray transients \cite{2015JHEAp...7..148K,2021ARA&A..59...21G}, but several TDEs have resulted in outflows or jets which shine in the radio bands \cite{2020SSRv..216...81A}. In the most common case in which no relativistic jet is launched (commonly referred to as ``non-jetted'' TDEs), the X-ray and optical/UV luminosity follows a roughly $t^{-5/3}$ decline, similar to the fallback rate. If the TDE leads to prompt disk formation, the X-rays are thought to arise from an accretion disk while the optical/UV emission arises from a large scale reprocessing layer \cite{2018ApJ...859L..20D,2022arXiv220606358C}. TDEs have also been observed to launch relativistic X-ray jets in a few cases. These jetted TDEs have been argued to arise due to a magnetically arrested disk \citep[MAD,][]{2003ApJ...589..444G} forming around the BH during the TDE \cite{2014MNRAS.437.2744T,2018ApJ...859L..20D,2019MNRAS.483..565C}, which leads to jet production via the Blandford-Znajek mechanism \cite{1977MNRAS.179..433B} extracting spin energy from the BH. Alternatively, jets may be produced thanks to radiative acceleration of gas through a narrow funnel region \cite{2020MNRAS.499.3158C}.

A handful of non-jetted TDEs have been observed to produce radio emission peaking at tens of GHz with $L_{\rm{radio}}\sim10^{37-39} \rm{erg\, s^{-1}}$. This emission is thought to arise from an outflow launched by the TDE with velocity $v\sim0.1c$ shocking on the gas surrounding the BH. Meanwhile, jetted TDEs produce bright radio emission peaking at $L_{\rm{radio}}\sim10^{40-42} \rm{erg\, s^{-1}}$. The appearance of radio emission is often delayed by several weeks from the initial appearance of the optical/UV/X-ray emission in non-jetted TDEs, which hints at some connection to the disk formation process to the occurrence of outflows.

After $\dot{M}_{\rm{fb}}$ rises to peak, it has previously been assumed that by this stage a circularized accretion disk has formed \cite{2018ApJ...859L..20D,2019MNRAS.483..565C}. The first direct demonstration of circularization near the peak fallback rate was recently demonstrated in a numerical simulation by \citet{Steinberg20221}. Multiple authors have argued in favor of a picture in which an inner accretion flow is surrounded by a quasi-spherical reprocessing layer since this naturally explains the sometimes delayed appearance of X-ray emission in optical/UV discovered TDEs \cite{2018ApJ...859L..20D,2022ApJ...937L..28T,2022arXiv220606358C}. This picture naturally arises if the accretion flow is actually super-Eddington \cite{2018ApJ...859L..20D}, which has motivated multiple studies of GRRMHD simulations magnetized of super-Eddington disks. Motivated by this fact and the demonstration that the ``standard and normal evolution'' (or SANE, citation) super-Eddington accretion disk also lead to viewing angle effects which may explain the behaviour in non-jetted TDEs \cite{2019MNRAS.483..565C}, we conducted a study of the outflows launched by SANE models in \citet{2022arXiv220606358C} and studied their radio-submm emission. We focus on these SANE models in this work as well.
 
The Next Generation Event Horizon Telescope (ngEHT) will provide more baseline coverage and faster response times than the previous mission. An estimate for ngEHT is to  double the antenna sites \cite{2019BAAS...51g.256D} of its 20 $\mu$as predecessor EHT, and thus the number of possible baseline pairs and triads of sites available for imaging jet/accretion flow/black hole systems would scale combinatorially (the number of baselines grows with the number of antennae as (N(N-1)/2). This could allow for interesting sources such as jets from nearby tidal disruption events to be imaged directly. In our previous work \cite{2022arXiv220606358C}, we provided the first demonstration that SANE super-Eddington accretion flows can produce radio emission which is bright enough at 230 GHz to be detected and resolved. Here we take things a step further and produce reconstructed images assuming such jets happen in the nearby universe. 

The detection rate of TDEs in the optical/UV/X-ray is expected to grow rapidly once the Large Synoptic Survey Telescope comes on line \cite{2019ApJ...873..111I,2020ApJ...890...73B}. Assuming rapid follow-up of TDEs in radio-submm bands finds detectable emission, this could provide a large number of targets for the ngEHT. As we demonstrated in \citet{2022arXiv220606358C}, some models produced detectable emission even at $\sim180$ Mpc. At this distance, a conservative estimate of the volume integrated TDE rate suggests more than 200 TDEs per year assuming volumetric TDE rates based on \citet{2016MNRAS.455..859S}. Even at $<40$ Mpc, we estimate that several TDEs should occur per year (see Figure 2 in \citealt{2022arXiv220606358C}) which suggests some nearby TDEs may become targets of opportunity during the ngEHT mission.

We stress that jets such as those in our first work on the subject of jets from SANE models of TDE accretion disks \cite{2022arXiv220606358C} do not resemble any previously detected radio TDEs. This may suggest most, or even all, TDEs do not form accretion disks which resemble SANE models to begin with. However, the number of TDEs that have appeared in the radio-submm in the first place is extremely small as of this writing with fewer than twenty radio TDEs reported. Furthermore, magnetic fields are certainly present in the forming disk, albeit dynamically subdominant to hydrodynamic effects early in the disk formation \cite{Sadowski2016,Curd2021}. Nevertheless, it is possible that after the disk circularizes, which \citet{Steinberg20221} suggests may take tens of days, the magnetic field builds up in a dynamo effect similar to \citet{Sadowski2016}. In this case, one would almost certainly expect the magnetic field to become dynamically important, in which case a magnetized outflow may be launched as in \citet{2022arXiv220606358C}. TDEs continue to surprise observers in terms of the range of behaviour, so such a jet formation channel may yet be discovered.



\section{Numerical Methods}

\subsection{GRRHMD Simulations}

Throughout this work, we often use gravitational units to define length and time. In particular, we use the gravitational radius
\begin{equation} \label{eq:rg}
    r_g = \dfrac{GM_{\rm{BH}}}{c^2}
\end{equation}
and the gravitational time
\begin{equation} \label{eq:tg}
    t_g = \dfrac{GM_{\rm{BH}}}{c^3},
\end{equation}
where $M_{\rm{BH}}$ is the mass of the black hole (BH). We also adopt the following definition for the Eddington mass accretion rate:
\begin{equation} \label{eq:mdotEdd}
  \dot{M}_{\rm{Edd}} = \dfrac{L_{\rm{Edd}}}{\eta_{\rm NT} c^2},
\end{equation}
where $L_{\rm{Edd}} = 1.25\times 10^{38}\, (M_{\rm{BH}}/M_\odot)\, {\rm erg\,s^{-1}}$ is the Eddington luminosity, and $\eta_{\rm{NT}}$ is the radiative efficiency of a Novikov-Thorne thin disk around a BH with spin parameter $a_*$ \cite{1973blho.conf..343N}.

We conduct an imaging analysis of GRRMHD simulations presented in \citet{2022arXiv220606358C}. In particular, we analyze the most massive BH models \texttt{m7a0.0-HR} and \texttt{m7a0.9-HR}, which are $M_{\rm{BH}}=10^7\, M_\odot$ BHs of spin $a_*=0$ and $0.9$ BHs. We specify the simulation diagnostics relevant for this work in Table \ref{tabKORALsims}. The simulations were conducted in 2D $(r,\vartheta)$ coordinates on a $N_r\times N_\vartheta = 640\times256$ grid with added resolution near the poles to adequately resolve both the disk and jet. The radial grid cells were logarithmically spaced with a maximum domain radius of $R_{\rm{max}}=10^5\,r_g$ to capture the large scale features of the jet.

\begin{table}[H] 
\centering
\caption{We tabulate the mass accretion rate $\dot{M}$, jet efficiency $\eta_{\rm{jet}}\equiv L_{\rm{jet}}/\dot{M}c^2$ (where $L_{\rm{jet}}$ is the jet power as defined in \cite{2022arXiv220606358C}), and total simulation duration $t_{\rm{sim}}$ for each \textsc{KORAL} simulation. Note that $\dot{M}$ and $\eta_{\rm{jet}}$ are time averaged over the final $50,000\,t_g$ of each simulation.\label{tabKORALsims}}
\begin{tabular}{lccc}
\toprule
\textbf{Model} & $\dot{M}$ & $\eta_{\rm{jet}}$ & $t_{\rm{sim}}$ \\
               & ($\dot{M}_{\rm{Edd}}$) & & $(t_g)$ \\
\midrule
\texttt{m7a0.0-HR}	& 12 & 0.24\% & $83,000$ \\
\texttt{m7a0.9-HR}	& 25 & 1.15\% & $81,200$ \\
\bottomrule
\end{tabular}
\end{table}

On horizon scales, gas is flowing across the BH horizon in an accretion disk due to angular momentum transport driven by the magneto rotational instability. The disk is optically thick and turbulent, and gas inside of the disk is advected with the gas across the BH horizon. However, an optically thin funnel above and below the disks exists. Here, radiation can escape freely and pushes on gas, accelerating a significant outflow. In addition, the funnel is magnetized and sometimes exhibits magnetization parameter $\sigma = b^2/\rho c^2 > 1$, where $b$ is the magnetic field strength and $\rho$ is the mass density of the plasma. In the jet, magnetic energy is partially converted into kinetic energy as it contributes to accelerating gas into an outflow.

\begin{figure}[H]
 \centering
 \includegraphics[width=\textwidth]{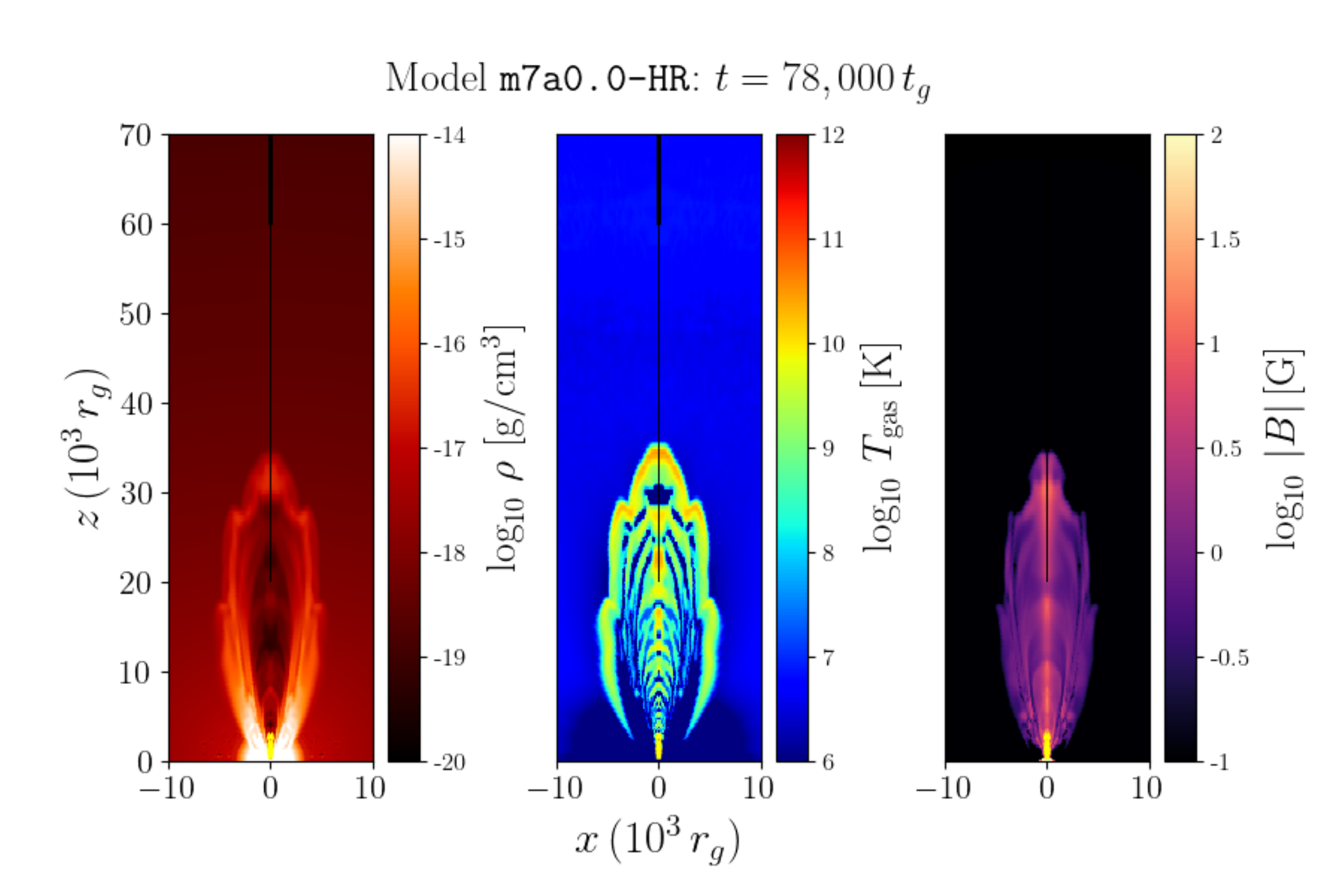}\\
 \includegraphics[width=\textwidth]{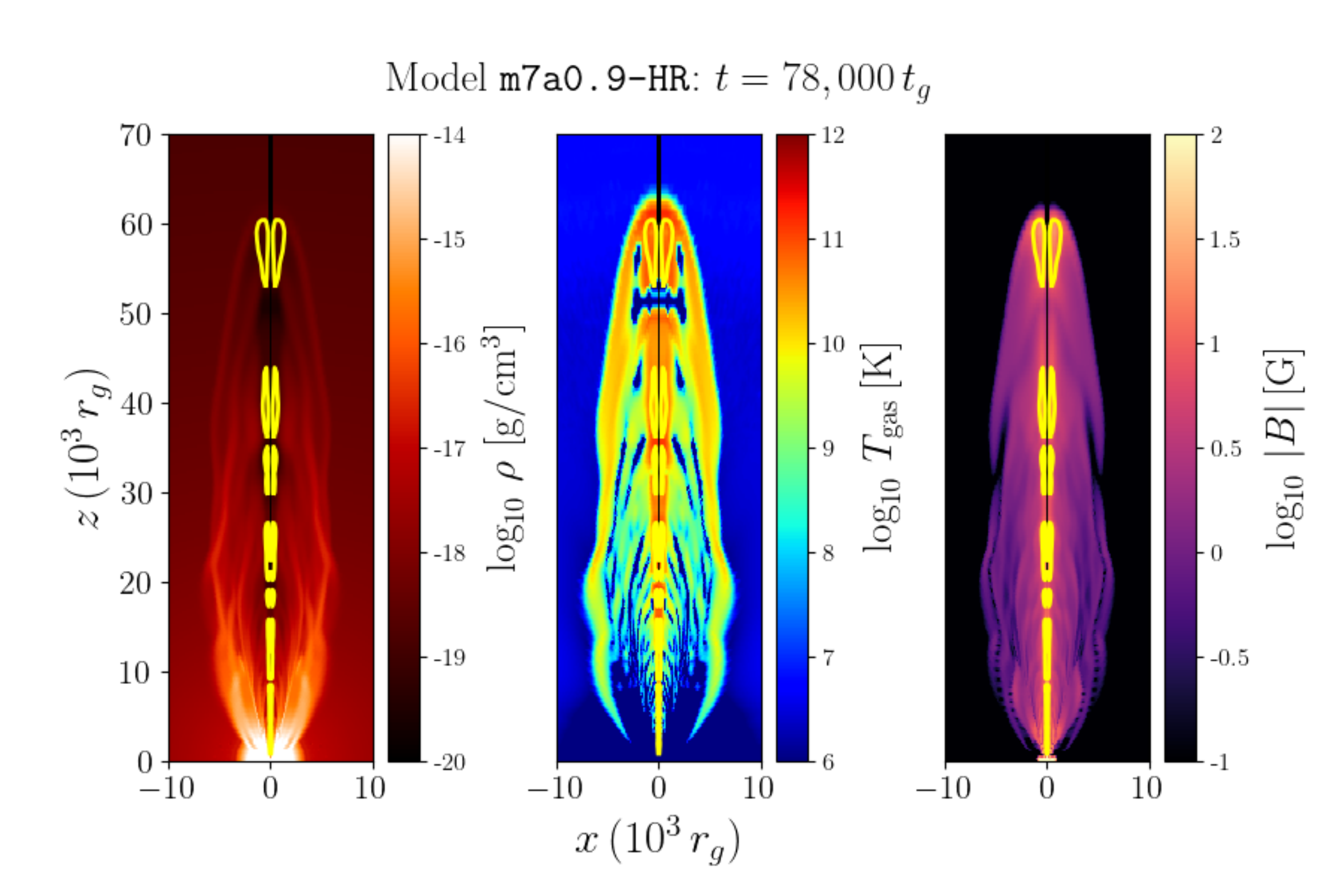}
 \caption{Here we show snapshots of the GRRMHD \textsc{KORAL} simulations that we post-process with \texttt{ipole}. All data is shown for $t=78,000\,t_g$ in both \texttt{m7a0.0-HR} (top) and \texttt{m7a0.9-HR} (bottom). The colors indicate the gas density $\rho$ (left), gas temperature $T_{\rm{gas}}$ (middle), and magnetic field strength $|B|$ (right) while the yellow contours indicate the $\sigma=1$ boundary, in which we set $\rho=0$ in the ray tracing step to prevent emission. \label{fig1}}
 \end{figure}

In both simulations, radiative and Poynting acceleration drive fast outflows. The jet reaches relativistic speeds with Lorentz factor $\gamma>5$ for model \texttt{m7a0.9-HR}, which is likely due to the BZ effect extracting spin energy from the BH, which may produce roughly $\sim1$ percent of the jet efficiency even though the magnetic flux threading the black hole is well below the MAD limit. The primary sites of dissipation are the jet head and internal shocks inside of the jet. Internal shocks are due to fast and slow moving gas interacting downstream of the jet head in addition to recollimation shocks. As we show in Figure \ref{fig1}, this results in a hot, magnetized jet which reaches large scales ($r>30,000\, r_g$) by the end of the simulation. The $a_*=0.9$ model has a significantly more magnetized jet and also produces a more powerful jet by a significant fraction (see Table \ref{tabKORALsims}).

\subsection{230 GHz Emission}

We post-process the KORAL simulation data with the general relativistic ray tracing (GRRT) code \texttt{ipole} \cite{Moscibrodzka2018,Yarza2020,Wong2022}, which includes synchrotron and Bremsstrahlung emission and absorption. The electron distribution function is assumed to be thermal. \citet{2019Galax...7...14O,2020MNRAS.493.5761O} demonstrated that large scale active galactic nuclei (AGN) jets can produce a two-temperature plasma. Motivated by their findings and the possibility that a two-temperature plasma will be produced due to shocks at the jet head and within the jet itself, we test a simple two-temperature jet model by scaling the electron temperature relative to the ion temperature via the plasma temperature ratio:
\begin{equation}
    \mathcal{R}=\dfrac{T_i}{T_e},
\end{equation}
where $T_i$ is the temperature of the ions, and $T_e$ is the temperature of the electrons, respectively. Note that $T_i$ is obtained directly from the \textsc{KORAL} simulation by setting $T_i=T_{\rm{gas}}$.

The peak of the radio-submm spectra in \texttt{m7a0.0-HR} is lower than that of \texttt{m7a0.9-HR}, so increasing $\mathcal{R}$ has a much more significant impact on the 230 GHz emission and can make the jet undetectable even at 10 Mpc for values of $\mathcal{R}>2$ \cite{2022arXiv220606358C}. It is possible that a non-thermal electron distribution will have greater high energy emission even as $\mathcal{R}$ increases, but we save an exploration of non-thermal electron models for a future analysis.

Each model was imaged at $230$ GHz. For both models, we image the simulation at times $t=38,000\,t_g$ and $t=78,000\,t_g$ for a difference in observing times of $\sim 23$ days. We choose a distance $D=10$ Mpc, $\mathcal{R}=1$, observing angle relative to the jet axis ($z$ in Figure \ref{fig1}) of $\theta=10,45,$ and $90^\circ$, respectively. Note that we use $\theta$ for the observer angle while $\vartheta$ is the polar angle in the \textsc{KORAL} grid coordinates. For model \texttt{m7a0.9-HR}, we also test limiting cases $D=100$ Mpc, $\mathcal{R}=1$ and $D=10$ Mpc, $\mathcal{R}=20$ imaged at $\theta=90^\circ$. The total 230 GHz flux of each ray traced model is tabulated in Table \ref{tab1}. 

We show a full library of each of the \texttt{ipole} images convolved with a Gaussian beam with a $20 \, \mu{\rm{as}}$ full width at half maximum (FWHM) in Figures \ref{figA1} and \ref{figA3}.

\begin{table}[H] 
\centering
\caption{Here we tabulate the 230 GHz flux density for each model given a specific time, viewing angle $\theta$, distance $D$, and temperature ratio $\mathcal{R}$.\label{tab1}}
\begin{tabular}{lcccccc}
\toprule
\textbf{Model} & \textbf{Time}  & \textbf{Distance}	&  $\mathcal{R}$ &	& $F_{230\,{\rm{GHz}}}$ & \\
               & $(t_g)$ & (Mpc) & & & (Jy) & \\
\midrule
  & & & & $\theta = 10^\circ$ & $\theta = 45^\circ$ & $\theta = 90^\circ$ \\
\cmidrule(lr){5-7}
\texttt{m7a0.0-HR}	& $38,000$ & $10$ & 1 & 0.219 & 0.214 & 0.074 \\
 	& $78,000$ & $10$ & 1 & 0.014 & 0.013 & 0.006 \\
\midrule
\texttt{m7a0.9-HR}	& $38,000$ & $10$ & 1 & 2.001 & 4.452 & 6.036 \\
 	& $78,000$ & $10$ & 1 & 11.968 & 26.780 & 35.092 \\
 	& & & & & & \\
	& $38,000$ & $10$ & 20 & - & - & 0.190 \\
 	& $78,000$ & $10$ & 20 & - & - & 0.485 \\
 	& & & & & & \\
	& $38,000$ & $100$ & 1 & - & - & 0.060 \\
 	& $78,000$ & $100$ & 1 & - & - & 0.351 \\
\bottomrule
\end{tabular}
\end{table}

\subsection{Synthetic ngEHT Observations and Image Reconstruction}

In order to test to what extent the jet features in our models can be observed, we simulated observations with a potential ngEHT array, consisting of the 2022 EHT stations and eleven additional stations, selected from \citet{Raymond2021} and similar to the ngEHT reference array used in the ngEHT Analysis Challenges \citep{Roelofs2022}. The new dishes were assumed to have a diameter of 10 m and a receiver temperature of 50 K, with the array operating at a bandwidth of 8 GHz. For each image, we simulated a 24-hour observation with a 50\% duty cycle with this array, using the {\tt ngehtsim}\footnote{\url{https://github.com/Smithsonian/ngehtsim}} library which makes use of {\tt eht-imaging} \citep{Chael2016, Chael2018} \citep[see also][]{Doeleman2022}. The atmospheric opacity was set to reflect a good day in April, using the top 1$\sigma$ quantile from the MERRA-2 data interpolated and integrated for each site on a 3-hour cadence for a 10-year period \citep{Gelaro2017, Paine2019}. Thermal noise was added to the complex visibilities, visibility phases were randomized, and no systematic visibility amplitude errors were added to the data. 

We subsequently used the regularized maximum likelihood framework in {\tt eht-imaging} to produce image reconstructions, with maximum entropy and (squared) total variation regularizers, fitting to visibility amplitudes and closure phases \citep[e.g.][]{Chael2016, Chael2018, EHT2019M87IV}. After establishing a set of well-performing imaging parameters on the {\tt m7a0.9-HR}, $\mathcal{R}=1$ model at $t_g=78,000$ and a distance of 10 Mpc (Fig. \ref{fig2}), we applied the same script to all other simulated datasets.

\section{Results}

In this section, we comment on the detectability of our models and then compare the \texttt{ipole} images with the reconstructed images. We comment on features which may be of interest in terms of the broader study of astrophysical jets.

\subsection{Reconstructed Images}

\begin{figure}[H]
 \centering
 \includegraphics[width=\textwidth]{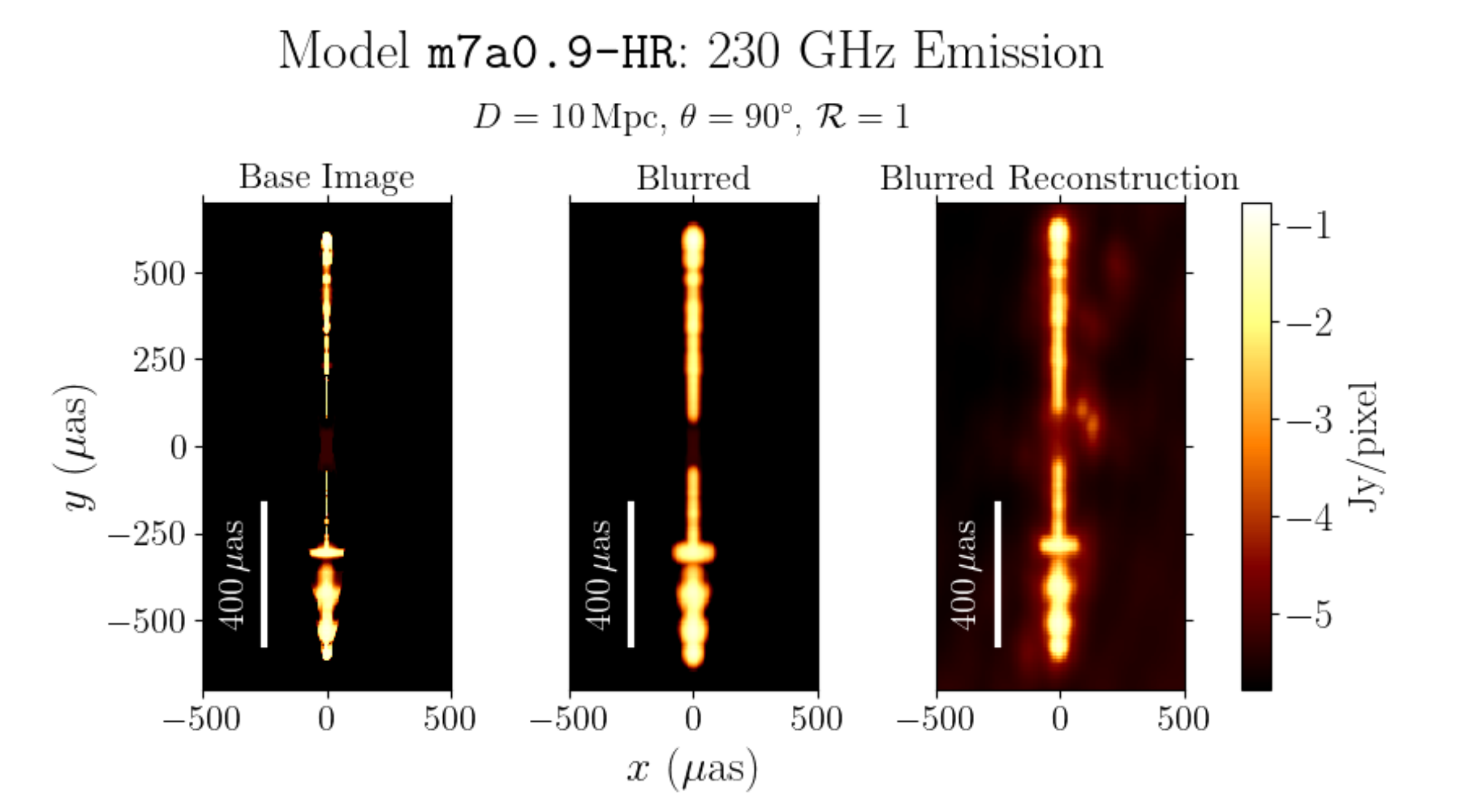}
 \caption{Model \texttt{m7a0.9-HR} at $t=78,000\, t_g$ imaged at $D=10$ Mpc with $\theta=90^\circ$ and $\mathcal{R}=1$. We show the base \texttt{ipole} image with no blurring (left), the base \texttt{ipole} image blurred via convolution with a $20\,\mu\rm{as}$ FWHM Gaussian beam (middle), and the reconstructed image blurred using the same Gaussian beam (right). \label{fig2}}
 \end{figure}

In the ray traced image (i.e. see the left panel in Figure \ref{fig2}), the jet head produces bright emission as it shocks on the circumnuclear medium (CNM). In addition, various shocks occur within the jet due to both slow/fast moving components colliding radially and due to recollimation shocks. This leads to dissipation within the jet and bright ``bubbles'' of emission at 230 GHz. As we show in the right panel of Figure \ref{fig2}, the jet head and the structures in the jet are faithfully reproduced in the reconstruction for favorable viewing angles ($\theta=45^\circ$ and $90^\circ$) as long as the source is nearby ($D=10$ Mpc).  Jets viewed near $\theta=10^\circ$ are dominated by emission from the jet head and distinguishing internal jet features would be unlikely. This can be seen by comparing the base images with the full library of reconstructed images for models \texttt{m7a0.0-HR} and \texttt{m7a0.9-HR} (Figures \ref{figA1}-\ref{figA4}).

\begin{figure}[H]
 \centering
 \includegraphics[width=\textwidth]{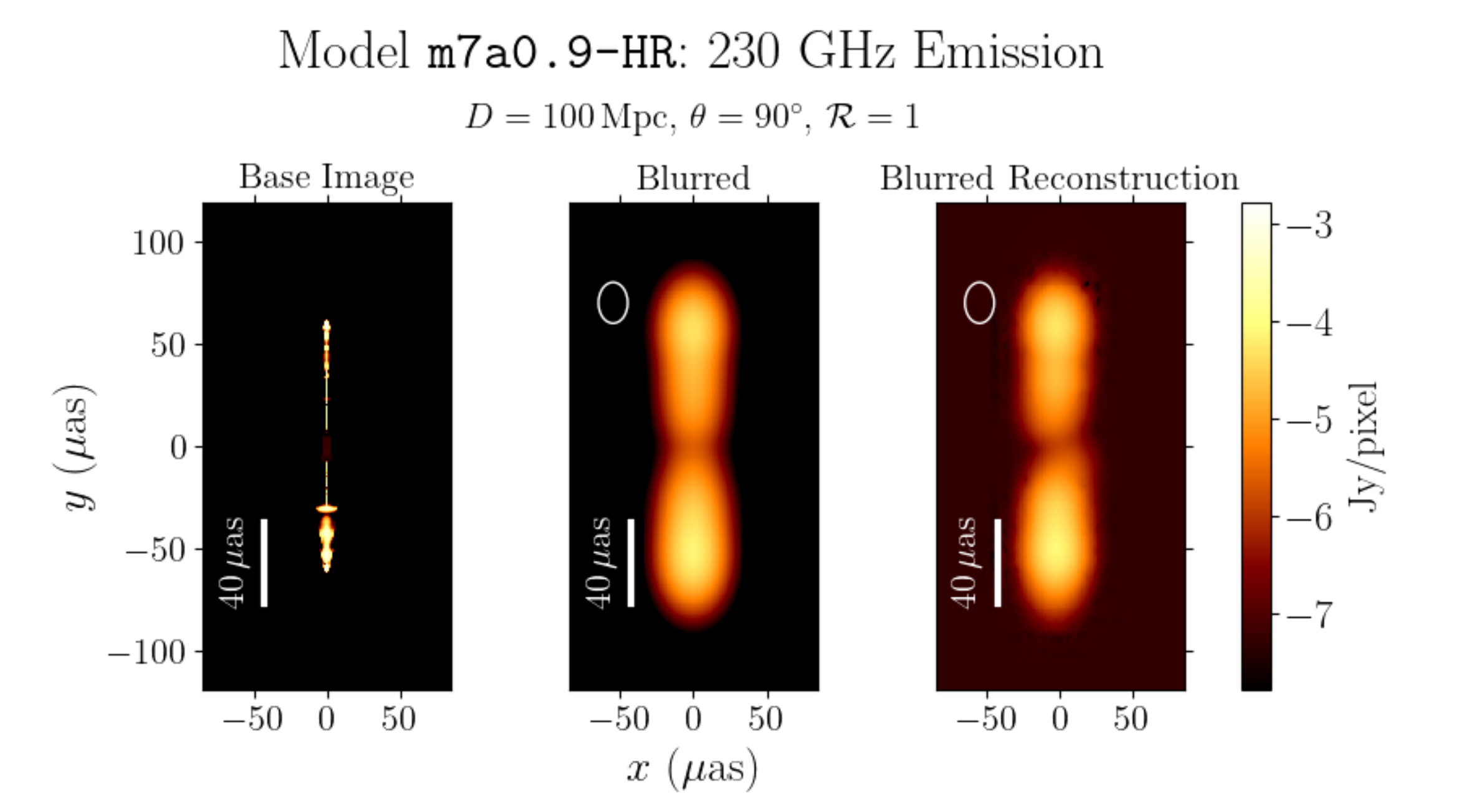}
 \caption{The same as Figure \ref{fig2} but for model \texttt{m7a0.9-HR} at $t=78,000\, t_g$ imaged at $D=100$ Mpc with $\theta=90^\circ$ and $\mathcal{R}=1$. \label{fig3}}
 \end{figure}

For distant sources ($D=100$ Mpc), distinguishing internal features is impossible and only the jet head can be fully distinguished in the reconstruction (right panel in Figure \ref{fig3}). This would still allow for the jet motion to be tracked, but detailed information is lost.

\subsection{Tracking Jet Motion}

In this subsection, we demonstrate that the original ray traced images and the reconstructed images allow for the jet motion to be tracked and yield similar results for the time evolution of the jet. The jet features are approximately Lorentzian, so we fit Lorentzian profiles to the image to find the position of the top and bottom jet in both the base images and the reconstructed images. We detail the peak finding algorithm in Appendix \ref{appB}. Since we cannot properly center the jet in the reconstructed images (there is no bright, central radiation from the near BH), we only measure the distance between the two jet peaks $y_1$ and $y_2$, respectively. We define the apparent jet length as:
\begin{equation}
    l_{\rm{jet}}=|y_2-y_1|.
\end{equation}
Note that we have not differentiated the 'top' or 'bottom' jet here as we are only concerned with the total distance between the jet heads. We obtain errors on the jet length from the error estimates of the jet head locations using standard error propagation analysis:
\begin{equation}
    \delta l_{\rm{jet}}=\sqrt{\delta y_1^2 + \delta y_2^2}.
\end{equation}

\begin{figure}[H]
 \centering
 \includegraphics[width=\textwidth]{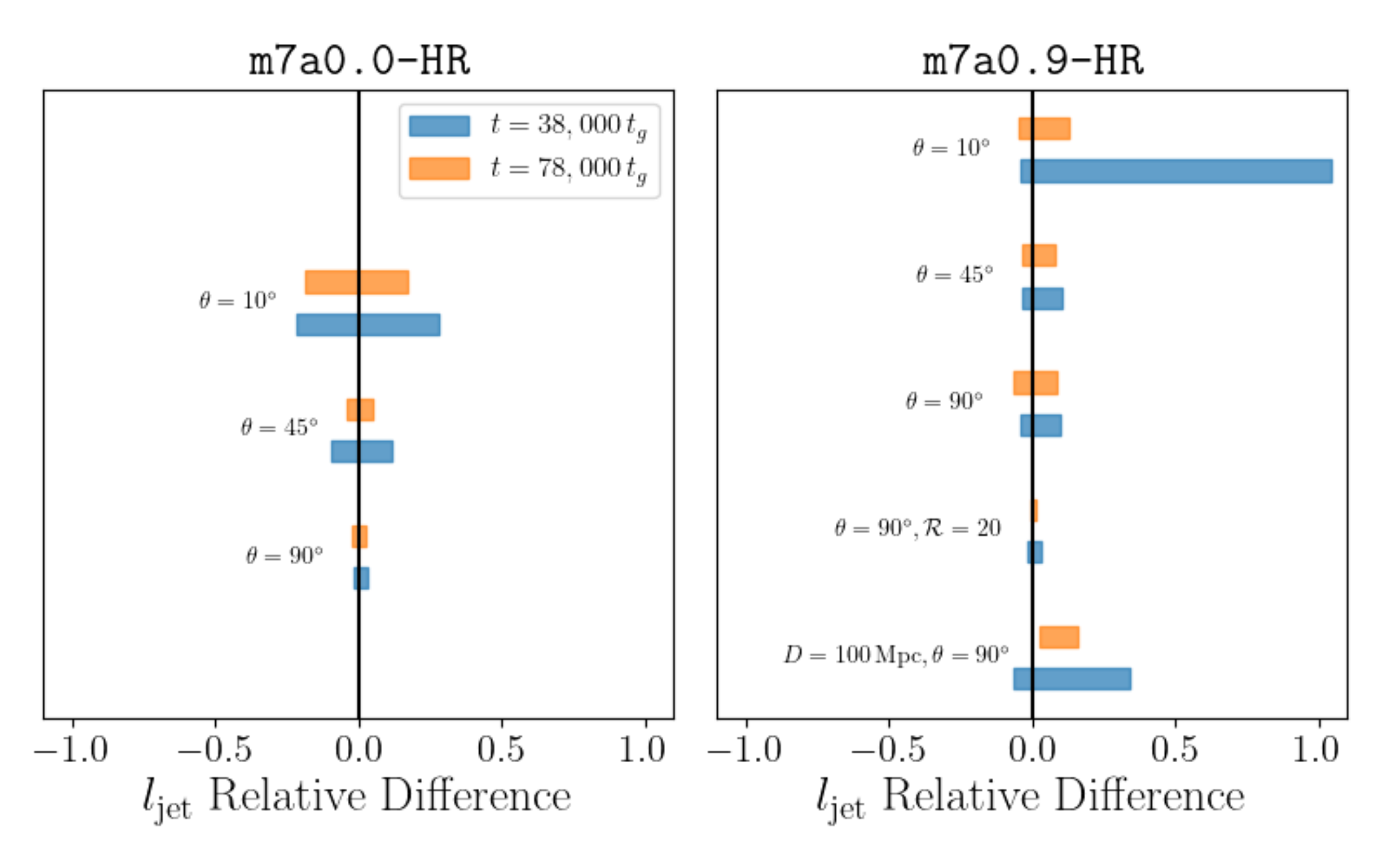}
 \caption{Here we show the relative difference at the 3 standard deviation level (colored horizontal bars) between the jet lengths obtained from the ray traced images $(l_{\rm{jet}})$ and the reconstructed images $(l_{\rm{jet,rec}})$. Models \texttt{m7a0.0-HR} (left panel) and \texttt{m7a0.9-HR} (right panel) are shown. For each choice of viewing angle $\theta$, distance to the source $D$, and $\mathcal{R}$, we show the data at $t=38,000\,t_g$ (blue) and $t=78,000\, r_g$. We indicate the viewing angle next to each pairing of error bars. For each model, except where explicitly indicated to be otherwise, we show $D=10$ Mpc and $\mathcal{R}=1$. The figure is described in the text. \label{fig4}}
 \end{figure}   

We compute the relative difference between the jet lengths for the ray traced ($l_{\rm{jet}}$) and reconstructed ($l_{\rm{jet,rec}}$) images in order to quantify the extent to which measurements of the jet length and velocity agree. We find that, generally, the ray traced and reconstructed images yield similar jet lengths within $3\delta l_{\rm{jet}}$ (Figure \ref{fig3}). In general, there are much larger errors on the fit for the Lorentzian profile's center at steep angles (i.e. see the relative difference for $\theta=10^\circ$). The agreement is also effected by how bright the source is, as illustrated by the shift to the right for model \texttt{m7a0.9-HR} when $D=100$ Mpc. We compile estimates of the jet lengths for each image in Table \ref{tab2}.

Since we lack centering information in the reconstructed images, we choose to estimate the jet velocity perpendicular to the line of sight by assuming both the top and bottom jet have the same speed. Then the velocity of the centroids of the jet in the source's frame are:
\begin{equation} \label{eq:vcen}
    v_{\rm{cen}}=\dfrac{1}{2}\dfrac{|l_{\rm{jet}}(t_2) - l_{\rm{jet}}(t_1)|}{t_2 - t_1}.
\end{equation}
Note that we use the same expression to derive the centroid velocity in the reconstructed images ($v_{\rm{cen,rec}}$) but replace $l_{\rm{jet}}$ with $l_{\rm{jet,rec}}$ in Equation \ref{eq:vcen}. Very Long Baseline Interferometry (VLBI) observations will provide the apparent motion of the jet, which may be superluminal due to relativistic effects. To account for this, we assume each centroid has a velocity of $v_{\rm{cen,rec}}$ and then estimate the apparent velocity ($v_{\rm{app,rec}}$) via the time in the observer's frame ($t_1'$,$t_2'$):
\begin{equation}
    v_{\rm{app,rec}}=\dfrac{1}{2}\dfrac{|l_{\rm{jet,rec}}(t_2') - l_{\rm{jet,rec}}(t_1')|}{t_2' - t_1'}=v_{\rm{cen,rec}}\left(1-\dfrac{v_{\rm{cen,rec}}}{c}\dfrac{\cos \theta}{\sin\theta}\right)^{-1}.
\end{equation}
Here we have used the relationship
\begin{equation}
  t_2'-t_1' = (t_2 - t_1)\left(1-\dfrac{v_{\rm{cen,rec}}}{c}\dfrac{\cos\theta}{\sin\theta}\right)
\end{equation} 
in the last expression. Note that the division by $\sin\theta$ is to account for the fact the $v_{\rm{cen,rec}}$ measures the velocity parallel to the line of sight with no time delay effects while we require an estimate of the velocity along the jet axis (which we can obtain since the geometry is fully known). We only present the apparent velocity for the reconstructed images ($v_{\rm{app,rec}}$) since this represents an estimate of what VLBI observations would truly see.

We use the data at $t_1=38,000\,t_g$ and $t_2=78,000\,t_g$ for each model to estimate the jet velocity. We tabulate the estimated centroid velocity and apparent velocity for each model in Table \ref{tab3}. We find excellent agreement between the velocities derived from the ray traced and reconstructed images. It is interesting to note the apparently faster jet for $\mathcal{R}=20$. We suspect the increased speed is due to the emitting material being dominated by material near the jet axis rather than some of the slower moving material around the jet head, which does not produce much emission at 230 GHz as the electron temperature is reduced. The more powerful jet model \texttt{m7a0.9-HR} demonstrates that such jets may appear as superluminal sources as we find a maximum $v_{\rm{app,rec}}\approx1.526c$ at $\theta=10^\circ$.

\begin{table}[H] 
\centering
\renewcommand{\arraystretch}{1.5}
\caption{Here we tabulate the estimated jet length for each model at each time for different choices of the distance $D$, viewing angle $\theta$, and plasma temperature ratio $\mathcal{R}$. We compare the jet length as computed from the base \texttt{ipole} image ($l_{\rm{jet}}$) and the reconstructed image ($l_{\rm{jet,rec}}$).\label{tab2}}
\begin{tabular}{lccclll}
\toprule
\textbf{Model} & \textbf{Time}  & \textbf{Distance}	& $\theta$ &  $\mathcal{R}$	& $l_{\rm{jet}}$ & $l_{\rm{jet,rec}}$ \\
               & $(t_g)$ & (Mpc) & & & $(r_g)$ & $(r_g)$ \\
\midrule
\texttt{m7a0.0-HR}	& $38,000$ & $10$ & $10^\circ$ & 1 & $5091^{+272}_{-272}$ & $4927^{+317}_{-317}$ \\
 	& $78,000$ & $10$ & $10^\circ$ & 1 & $9042^{+356}_{-356}$ & $9104^{+407}_{-407}$ \\
 & & & & &\\
 	& $38,000$ & $10$ & $45^\circ$ & 1 & $22,006^{+123}_{-123}$ & $21,774^{+731}_{-731}$\\
 	& $78,000$ & $10$ & $45^\circ$ & 1 & $39,832^{+318}_{-318}$ & $39,678^{+501}_{-501}$\\
 & & & & &\\
 	& $38,000$ & $10$ & $90^\circ$ & 1 & $28,658^{+148}$ & $28,463^{+173}_{-173}$\\
 	& $78,000$ & $10$ & $90^\circ$ & 1 & $57,068^{+401}_{-401}$ & $56,844^{+237}_{-237}$\\
\midrule
\texttt{m7a0.9-HR}	& $38,000$ & $10$ & $10^\circ$ & 1 & $8457^{+446}_{-446}$ & $4191^{+726}_{-726}$\\
 	& $78,000$ & $10$ & $10^\circ$ & 1 & $17,585^{+233}_{-233}$ & $16,837^{+726}_{-726}$\\
 & & & & &\\
 	& $38,000$ & $10$ & $45^\circ$ & 1 & $33,128^{+336}_{-336}$ & $31,941^{+683}_{-683}$\\
 	& $78,000$ & $10$ & $45^\circ$ & 1 & $73,535^{+890}_{-890}$ & $71,855^{+1058}_{-1058}$\\
 & & & & &\\
 	& $38,000$ & $10$ & $90^\circ$ & 1 & $48,468^{+306}_{-306}$ & $46,972^{+1028}_{-1028}$\\
 	& $78,000$ & $10$ & $90^\circ$ & 1 & $111,503^{+1085}_{-1085}$ & $110,327^{+2657}_{-2657}$\\
 & & & & &\\
 	& $38,000$ & $10$ & $90^\circ$ & 20 & $49,457^{+150}_{-150}$ & $48,930^{+358}_{-358}$\\
 	& $78,000$ & $10$ & $90^\circ$ & 20 & $119,897^{+173}_{-173}$ & $119,256^{+415}_{-415}$\\
 & & & & &\\
 	& $38,000$ & $100$ & $90^\circ$ & 1 & $46,715^{+400}_{-400}$ & $40,244^{+2714}_{-2714}$\\
 	& $78,000$ & $100$ & $90^\circ$ & 1 & $109,091^{+959}_{-959}$ & $98,822^{+2075}_{-2075}$\\
\bottomrule
\end{tabular}
\end{table}

\begin{table}[H] 
\centering
\renewcommand{\arraystretch}{1.5}
\caption{Here we tabulate the estimated jet velocity for each model for different choices of the distance $D$, viewing angle $\theta$, and plasma temperature ratio $\mathcal{R}$. The velocities shown are calculated using the base \texttt{ipole} images ($v_{\rm{cen}}$), the reconstructed images ($v_{\rm{cen,rec}}$), and the reonstructed images and accounting for the possibility of superluminal motion ($v_{\rm{app,rec}}$). \label{tab3}}
\begin{tabular}{lcccccc}
\toprule
\textbf{Model} & \textbf{Distance}	& $\theta$ &  $\mathcal{R}$	& $v_{\rm{cen}}$  &  $v_{\rm{cen,rec}}$ &  $v_{\rm{app,rec}}$\\           
 & (Mpc) & & & $(c)$ & $(c)$ & $(c)$\\
\midrule
\texttt{m7a0.0-HR} & $10$ & $10^\circ$ & 1 & $0.049^{+0.006}_{-0.006}$ & $0.052^{+0.006}_{-0.006}$ & $0.074^{+0.009}_{-0.009}$\\
 	& $10$ & $45^\circ$ & 1 & $0.223^{+0.005}_{-0.005}$ & $0.224^{+0.011}_{-0.011}$ & $0.288^{+0.014}_{-0.014}$ \\
 	& $10$ & $90^\circ$ & 1 & $0.355^{+0.005}_{-0.005}$ & $0.355^{+0.004}_{-0.004}$ & $0.355^{+0.004}_{-0.004}$  \\
\midrule
\texttt{m7a0.9-HR} & $10$ & $10^\circ$ & 1 &  $0.114^{+0.006}_{-0.006}$ & $0.158^{+0.011}_{-0.011}$ & $1.526^{+0.102}_{-0.102}$  \\
 	& $10$ & $45^\circ$ & 1 & $0.505^{+0.012}_{-0.012}$ & $0.499^{+0.016}_{-0.016}$ & $0.996^{+0.031}_{-0.031}$  \\
 	& $10$ & $90^\circ$ & 1 & $0.788^{+0.014}_{-0.014}$ & $0.792^{+0.036}_{-0.036}$ & $0.792^{+0.036}_{-0.036}$  \\
 	& $10$ & $90^\circ$ & 20 & $0.88^{+0.002}_{-0.002}$ & $0.879^{+0.007}_{-0.007}$ & $0.879^{+0.007}_{-0.007}$  \\
 	& $100$ & $90^\circ$ & 1 & $0.780^{0.013}_{-0.013}$ & $0.732^{+0.043}_{-0.043}$ & $0.732^{+0.043}_{-0.043}$  \\
\bottomrule
\end{tabular}
\end{table}

\section{Discussion}

\subsection{Extracting Jet Physics from VLBI Images}

A key feature of the jets in our models is the bright ``bubbles'' (or knots) of 230 GHz emission, which appear to correlate with recollimation shocks. Such structures have been seen in VLBI images of various AGN jets \cite{2005AJ....130.1418J,2013AJ....146..120L,2014ApJ...787..151C}. Previous simulations of jets in various astrophysical contexts have demonstrated that recollimation occurs when there is a pressure mismatch between the jet and the surrounding medium, which could be a static atmosphere or a slower moving jet sheath \cite{2012MNRAS.422.2282K,2012ApJ...750...68L,2015ApJ...809...38M,2017A&A...606A.103H}. The number of recollimation shocks along the jet axis are dependent on the properties of the jet and medium. It is therefore possible that direct VLBI of TDE jets will allow in depth modeling of jet launching and could also aid in constraining the properties of the surrounding medium. For instance, one work successfully applied simulations of MHD jets to constrain properties of BL Lacartae, which is a blazar jet with recollimation features \cite{2016ApJ...817...96G}.

We suggest that a similar approach may be applied in TDE jets. With a suitable exploration of the parameter space, it is conceivable that an analysis similar to that of \cite{2016ApJ...817...96G} could be applied to TDE jets in cases where VLBI is possible. A broader exploration of TDE jets through various simulation methodologies is strongly suggested. We plan to explore the effects of the ambient medium, magnetic field strength, and disk accretion rate on the jet properties in the case of a SANE, super-Eddington disk in a future work.

\subsection{Proposed Observational Methodology}

Our synthetic ngEHT observations demonstrate that a 24-hour observation may be sufficient to study both the structure and/or motion of newly born TDE jets. This is much shorter than the fallback time, which is on the order of month(s), as well as the duration of radio emission in several TDEs, which can sometimes be visible for years \cite{2020SSRv..216...81A}. Our suggested observational strategy is conducting rapid followup of newly discovered optical/X-ray TDEs when they are near the peak of their emission in order to study both the early- and late-time properties of their jets (if present). Observations with a single telescope at 230 GHz can be conducted to search for TDEs emitting in the radio. If emission at 230 GHz is detected, we suggest that the ngEHT conduct VLBI follow-up of targets within no more than a month. Our imaging simulations were done assuming a full ngEHT array consisting of the 2022 EHT stations plus 11 additional sites, but depending on the target not all sites may need to be available in order to obtain a high-fidelity image reconstruction.

Unlike many other EHT/ngEHT targets, TDEs will appear randomly across the sky and the ngEHT will need to be capable of follow-up observations on the order of a week to weeks. Our current modeling of jets and outflows from super-Eddington disks is too sparse to make predictions regarding how long the jets will be visible at 230 GHz. However, if radio TDE observations are any indicator, emission may persist for many months \cite{2020SSRv..216...81A}. 

TDEs provide an excellent laboratory for studying jet/accretion/black hole systems across a wide range of accretion states over a relatively short period of time ($\sim$1-a few years). In several cases, TDEs have shown state transitions after several hundred days in the X-ray which are likely associated with the evolution of the disk as the mass accretion rate declines. \citet{2016MNRAS.455..859S} argue for instance that the transition from a thick, super-Eddington disk to a thin disk can explain the jet shut-off in jetted TDEs such as \textit{Swift} J1644+57 and \textit{Swift} J2058+05 \cite{2013ApJ...767..152Z,2015ApJ...805...68P}, but recent simulations \cite{2022arXiv220912081C,2022ApJ...935L...1L} demonstrate that MAD is possible even for thinner accretion disks. As such, long term VLBI monitoring is strongly suggested as this would allow for (1) the radio-submm emission of the outflows to be characterized and compared to the behaviour of the accretion flow, and (2) the direct study of how the jet evolves morphologically as the disk state changes.

Another attractive potential target which we have yet to attribute a self-contained study to is jetted TDEs. These TDEs are extremely rare and current observations suggest only about 1\% of all TDEs will produce powerful relativistic jets. These jets will produce extremely bright radiation in the X-ray as well as the radio-submm. However, most have been distant due to the lower probability of their occurrence. Should a jetted TDE occur nearby enough for VLBI to resolve the jet, we strongly suggest such jets be treated as targets of opportunity for the ngEHT.

Lastly, a recent TDE AT2018hyz showed a late outflow ($\sim3$ years after the initial outburst) and brightened in the radio over several hundred days \cite{2022ApJ...938...28C}. Unlike many other radio TDEs, relatively bright 240 GHz emission was detected. If AT2018hyz is in fact a jet instead of a spherical outflow, \citet{2022ApJ...938...28C} estimate that the velocity could reach $\lesssim0.6c$. AT2018hyz is a relatively nearby TDE at $\sim 204$ Mpc, but the flux density at the time of detection ($\sim0.2$ mJy at 240 GHz) makes it too dim for ngEHT follow-up. However, placing AT2018hyz at $\sim50$ Mpc would shift the flux density to $\sim5$ mJy, which is the minimum estimated flux density required for an ngEHT VLBI detection. Future TDEs will likely be monitored across the radio-submm, so nearby targets of opportunity such as late radio TDEs like AT2018hyz should be considered should they show significant radio emission.

\section{Conclusions}

In this work, we have demonstrated through a synthetic imaging analysis that TDE jets resembling the GRRMHD models presented in \cite{2022arXiv220606358C} are compelling ngEHT targets. We also confirm that the detection limits considered in \cite{2022arXiv220606358C} are roughly applicable as \texttt{m7a0.0-HR} did not produce detectable emission at a distance of 100 Mpc in our imaging analysis. 

Various shock features in the jet are visible for the $10$ Mpc images we consider, and studying the jet morphology in these cases could aid in characterizing the environment of the BH. Most TDEs that occur during the ngEHT mission will be farther away, but the apparent motion, which may be superluminal, can be extracted in such cases.

We suggest that the ngEHT be utilized for radio follow-up of TDEs. Our models study the birth of a TDE jet in the first $\sim48$ days after the disk forms under the assumption that the disk is SANE, super-Eddington, and threaded by a dynamically important magnetic field. However, TDEs which occur nearby and have jet properties similar to jetted TDEs such as \textit{Swift} J1644+57 or AT2018hyz may provide interesting targets of opportunity.


\funding{
Brandon Curd was supported by NSF grant AST-1816420, and made use of computational support from NSF via XSEDE/ACCESS resources (grant TG-AST080026N). Razieh Emami acknowledges the support by the Institute for Theory and Computation at the Center for Astrophysics as well as grant numbers 21-atp21-0077, NSF AST-1816420 and HST-GO-16173.001-A for very generous supports. Freek Roelofs was supported by NSF grants AST-1935980 and AST-2034306. This work was supported by the Black Hole Initiative at Harvard University, made possible through the support of grants from the Gordon and Betty Moore Foundation and the John Templeton Foundation. The opinions expressed in this publication are those of the author(s) and do not necessarily reflect the views of the Moore or Templeton Foundations.
}

\textbf{Acknowledgements}
\qquad   We graciously acknowledge Koushik Chatterjee, Lani Oramas, Joaquin Duran, and Hayley West for fruitful discussions helpful in the preparation of this work.
\appendixtitles{no} 
\appendixstart
\appendix

\section[\appendixname~\thesection]{Full Image Library}

\begin{figure}[H]
 \centering
 \includegraphics[width=\textwidth]{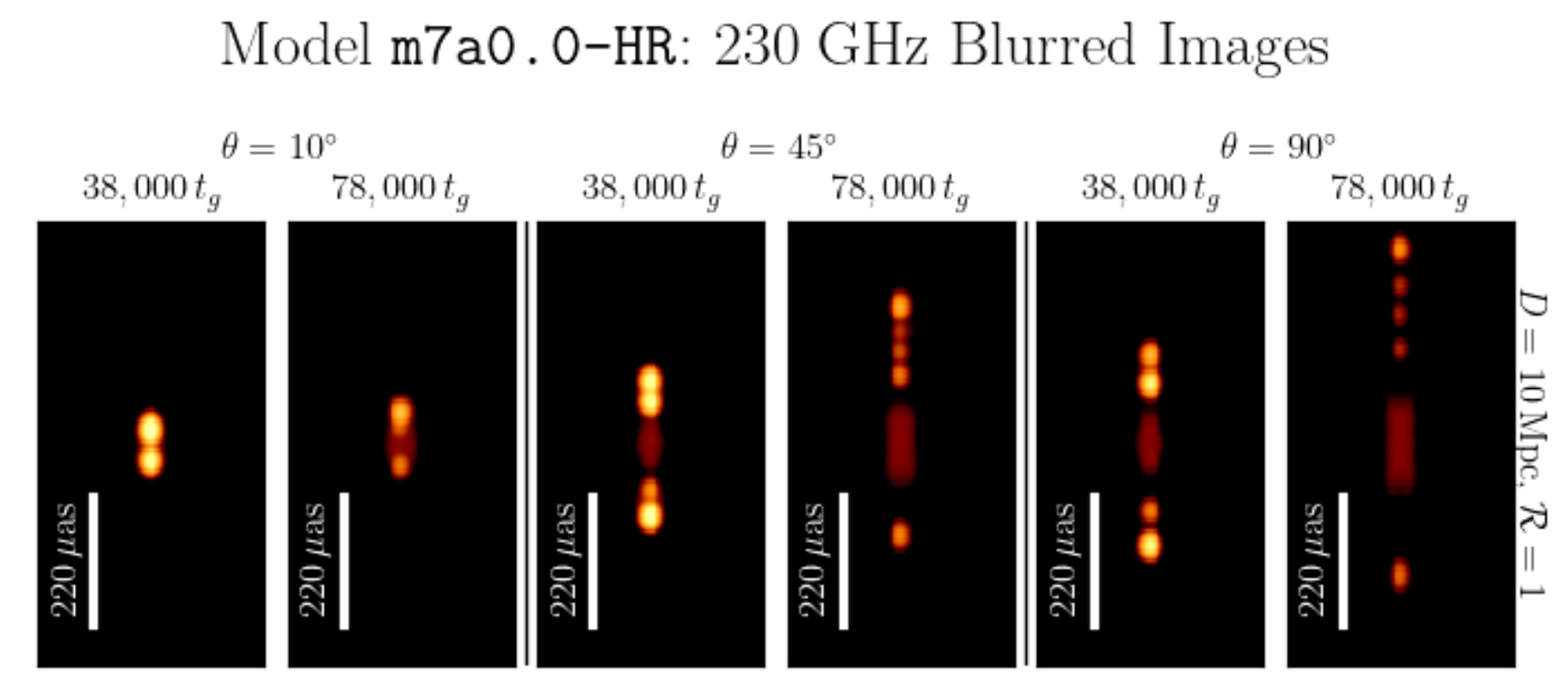}
 \caption{Here we show the full library of base \texttt{ipole} images for model \texttt{m7a0.0-HR}. The time of each column is indicated at the top while the distance $D$ and plasma temperature ratio $\mathcal{R}$ for each row are indicated on the right. The angle of the observer relative to the jet axis is indicated above each set of two rows. Each image spans $550\times770\, \mu\rm{as}^2$ and is blurred by convolving the base image with a Gaussian beam with a FWHM of $20\, \mu\rm{as}$. The colour scale is logarithmic, spanning three orders of magnitude, and each image uses the same maximum for the intensity scale. \label{figA1}}
 \end{figure}

\begin{figure}[H]
 \centering
 \includegraphics[width=\textwidth]{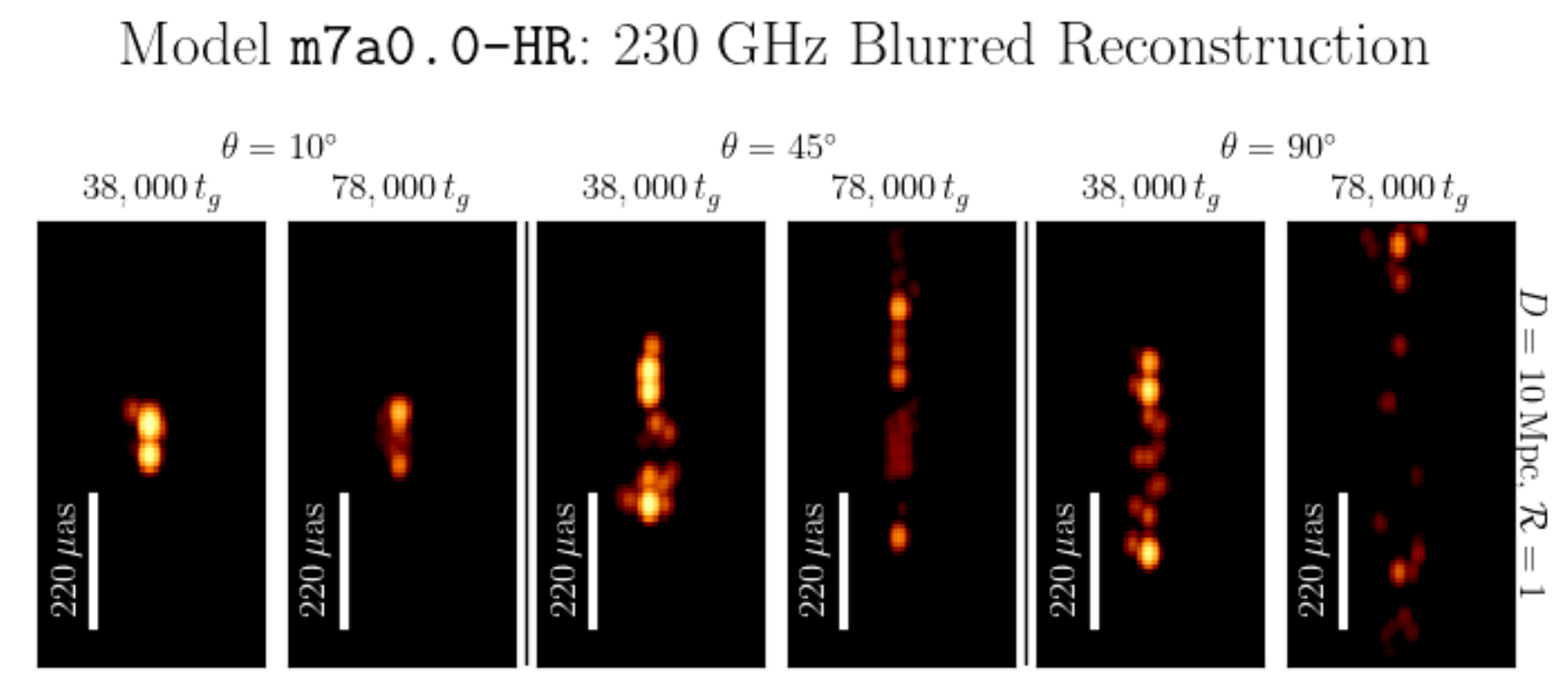}
 \caption{The same as Figure \ref{figA1} but showing the reconstructed images. Note that the intensity scale for each panel is the same as the corresponding panel in Figure \ref{figA1} for comparison. \label{figA2}}
 \end{figure}

\begin{figure}[H]
 \centering
 \includegraphics[width=\textwidth]{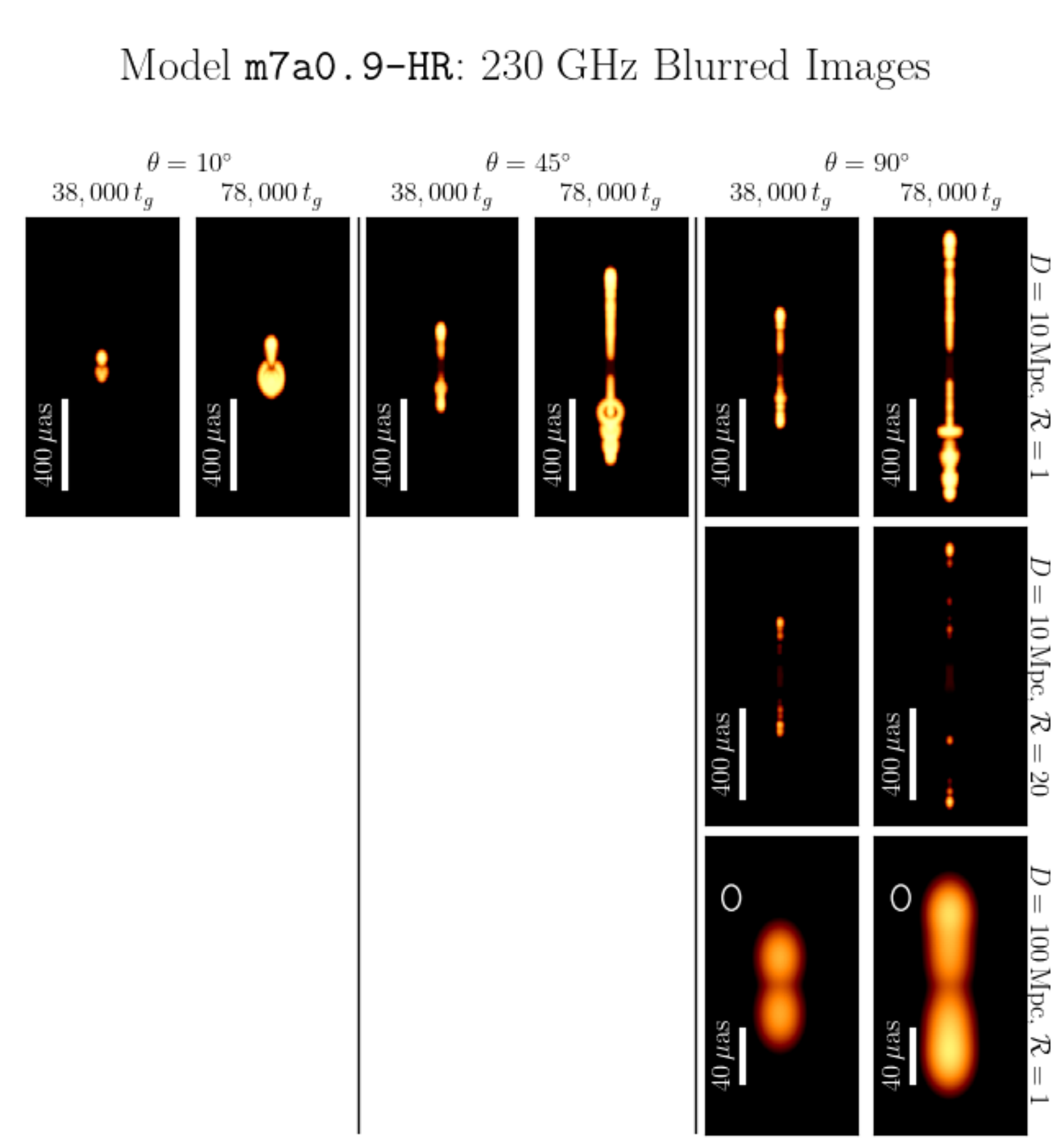}
 \caption{Here we show the full library of base \texttt{ipole} images for model \texttt{m7a0.9-HR}. Each image in the top and middle rows spans $1000\times1400\, \mu\rm{as}^2$ while the bottom row spans $170\times238\, \mu\rm{as}^2$. Each image is blurred by convolving the base image with a Gaussian beam with a FWHM of $20\, \mu\rm{as}$ (indicated by the white circle in the bottom right panel). The time of each column is indicated at the top while the distance $D$ and plasma temperature ratio $\mathcal{R}$ for each row are indicated on the right. The angle of the observer relative to the jet axis is indicated above each set of two rows. The colour scale is logarithmic, spanning three orders of magnitude in each image. We use the same colour scale for the $D=10$ Mpc images, but reduce the maximum by an order of magnitude in the $D=100$ Mpc images to better show the image features. \label{figA3}}
 \end{figure}

\begin{figure}[H]
 \centering
 \includegraphics[width=\textwidth]{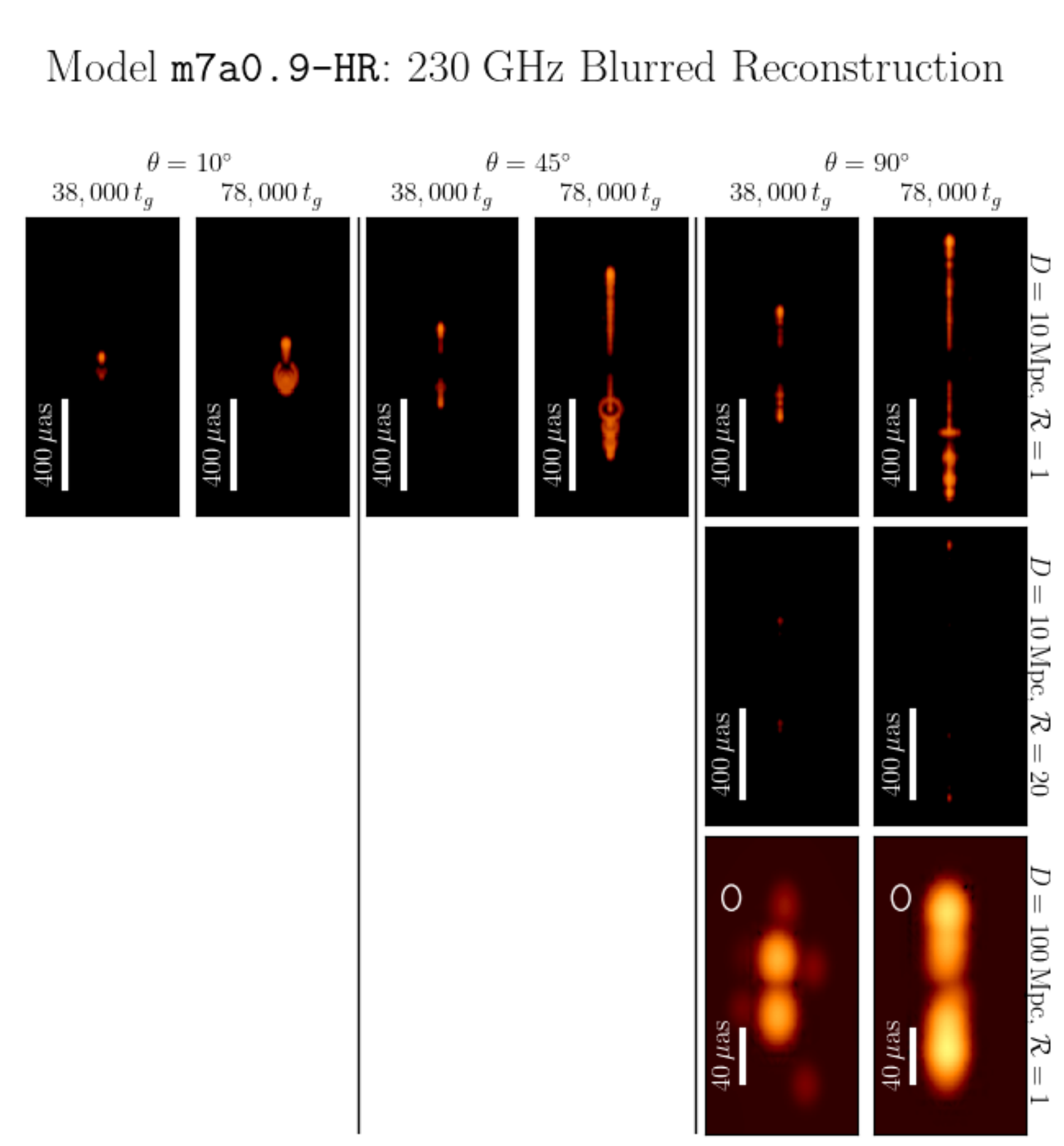}
 \caption{The same as Figure \ref{figA3} but showing the reconstructed images. Note that the intensity scale for each panel is the same as the corresponding panel in Figure \ref{figA3} for comparison. \label{figA4}}
 \end{figure}

In Figures \ref{figA1}-\ref{figA4}, we show the full library of images analyzed in this work. All of the base images were ray traced at $\nu=230$ GHz and then convolved with a Gaussian beam with a FWHM of 20 $\mu$as. Similarly, we blur the reconstructed image using the same beam for comparison. Note that in Figures \ref{figA2} and \ref{figA4}, we have also shifted the reconstructed image to be approximately centered for comparison with the base image. In general, bright features are represented quite well in the reconstruction; however, some noise is introduced in dimmer sources (i.e. see the reconstruction of \texttt{m7a0.9-HR} at $D=100$ Mpc).

\section[\appendixname~\thesection]{Fitting Procedure for Jet Head Position} \label{appB}

Here we describe the algorithm implemented to estimate the position of the top and bottom jet heads in each image. In order to track the jet's motion, we first smooth the data (either the base image or the reconstructed image) with a Gaussian beam which assumes an angular resolution for the VLBI observations of $\Delta \theta =20 \mu{\rm{as}}$. We then bin the data along the symmetry axis of the jet ($y$) by summing the flux along each row ($x$). 

Since the images appear to be roughly Lorentzian, we first attempt to fit a double Lorentzian function of the form:
\begin{equation} \label{eq:lord}
    f_d(y) = a_1\dfrac{w_1^2}{(y-y_{{\rm{cen}},2})^2 + w_1^2} + a_2\dfrac{w_2^2}{(y-y_{{\rm{cen}},2})^2 + w_2^2},
\end{equation}
where $a$ is the amplitude, $w$ is the width, and $y_{\rm{cen}}$ is the center defining the curve. We take $y_{\rm{cen}}$ as a measure of the jet head location. If this fitting procedure does not produce a good fit, we find the peaks by performing a two-step fitting procedure in which we fit a single Lorentzian:
\begin{equation} \label{eq:lor1}
    f_1(y) = a_1\dfrac{w_1^2}{{(y-y_{{\rm{cen}},1}})^2+w_1^2},
\end{equation}
and then subtract the fit $f_1(y)$ from the data and then fit the second peak with:
\begin{equation} \label{eq:lor2}
    f_2(y) = a_2\dfrac{w_2^2}{{(y-y_{{\rm{cen}},2}})^2+w_2^2}.
\end{equation}

\begin{figure}[H]
 \centering
 \includegraphics[width=0.49\textwidth]{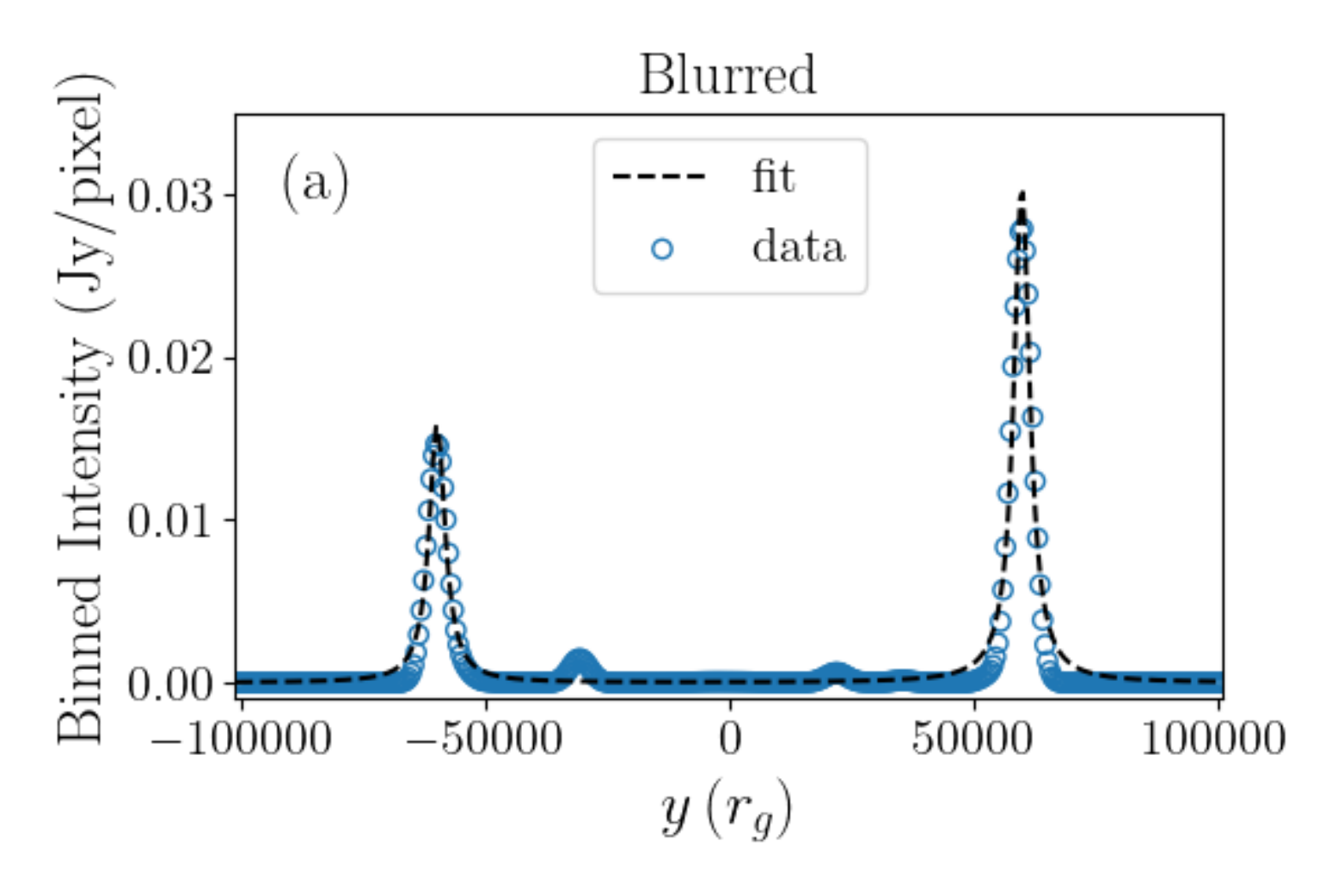}
 \includegraphics[width=0.49\textwidth]{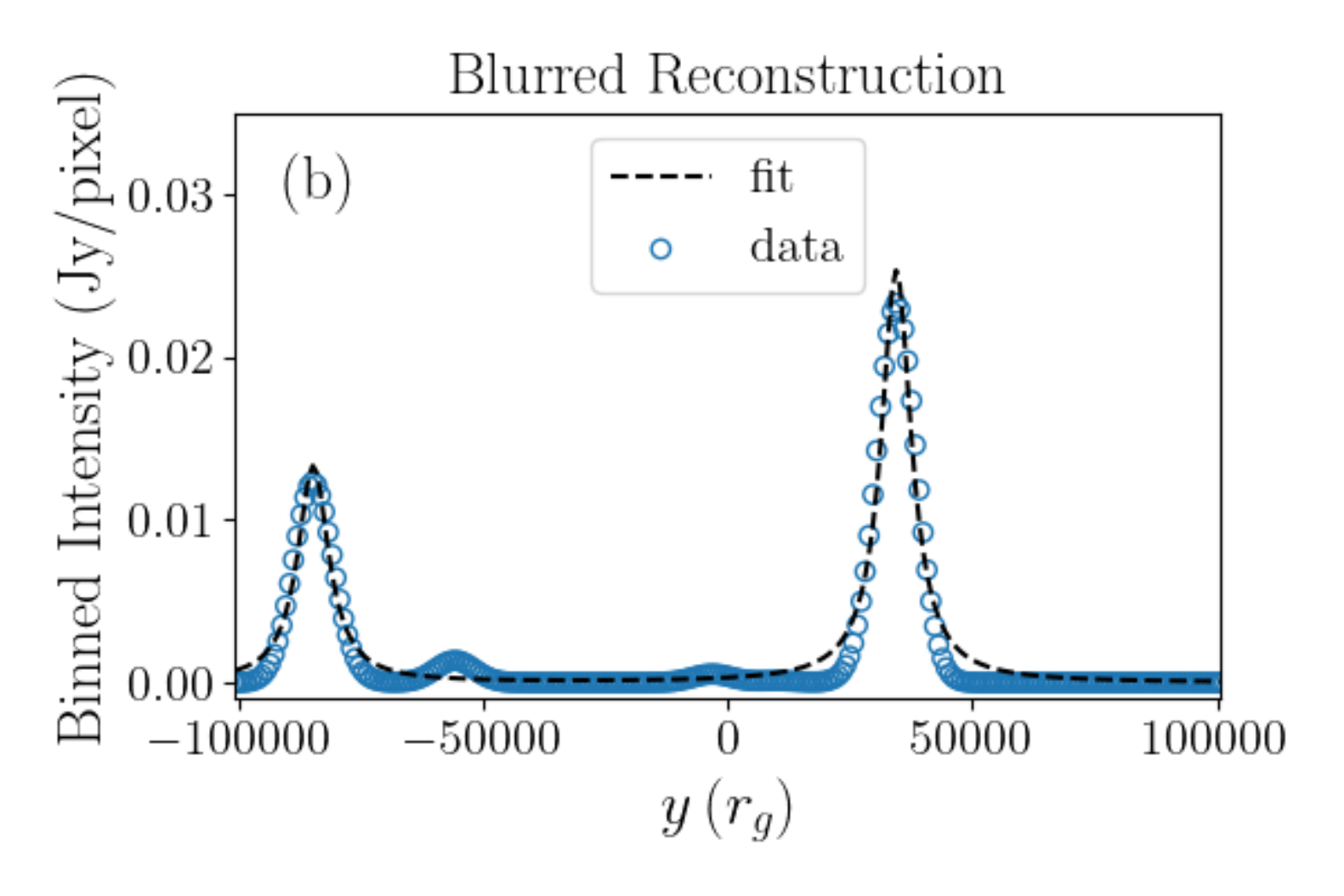}
 \caption{We demonstrate the fit performance using a ray traced image of model \texttt{m7a0.9-HR} at $t=78,000\,t_g$, $\theta=90^\circ$, $D=10$ Mpc, and $\mathcal{R}=20$. We show the $x$-binned data and a double Lorentzian (Equation \ref{eq:lord}) fit for (a) the blurred base image, and (b) the blurred reconstruction. \label{figA5}}
 \end{figure}

We implement the \texttt{python} package, \texttt{SciPy} \cite{2020SciPy-NMeth}, to optimize the curve(s) and estimate the jet head positions and errors. We show an example of the data and the Lorentzian fit for both the base image and the Reconstruction in Figure \ref{figA5}. The reconstruction tends to be a bit broader, but the fitting procedure works equally well for all images and reconstructed images.

\begin{adjustwidth}{-\extralength}{0cm}

\reftitle{References}

\bibliography{main}

\begin{thebibliography}{999}

\bibitem[{Hills}(1975)]{1975Natur.254..295H}
{Hills}, J.G.
\newblock {Possible power source of Seyfert galaxies and QSOs}.
\newblock {\em Nature} {\bf 1975}, {\em 254},~295--298.
\newblock {\url{https://doi.org/10.1038/254295a0}}.

\bibitem[{Rees}(1988)]{1988Natur.333..523R}
{Rees}, M.J.
\newblock {Tidal disruption of stars by black holes of {}10$^{6}$-{}10$^{8}$
  solar masses in nearby galaxies}.
\newblock {\em Nature} {\bf 1988}, {\em 333},~523--528.
\newblock {\url{https://doi.org/10.1038/333523a0}}.

\bibitem[{Guillochon} and {Ramirez-Ruiz}(2013)]{2013ApJ...767...25G}
{Guillochon}, J.; {Ramirez-Ruiz}, E.
\newblock {Hydrodynamical Simulations to Determine the Feeding Rate of Black
  Holes by the Tidal Disruption of Stars: The Importance of the Impact
  Parameter and Stellar Structure}.
\newblock {\em ApJ} {\bf 2013}, {\em 767},~25,
  \href{http://xxx.lanl.gov/abs/1206.2350}{{\normalfont
  [arXiv:astro-ph.HE/1206.2350]}}.
\newblock {\url{https://doi.org/10.1088/0004-637X/767/1/25}}.

\bibitem[{Mainetti} \em{et~al.}(2017){Mainetti}, {Lupi}, {Campana}, {Colpi},
  {Coughlin}, {Guillochon}, and {Ramirez-Ruiz}]{2017A&A...600A.124M}
{Mainetti}, D.; {Lupi}, A.; {Campana}, S.; {Colpi}, M.; {Coughlin}, E.R.;
  {Guillochon}, J.; {Ramirez-Ruiz}, E.
\newblock {The fine line between total and partial tidal disruption events}.
\newblock {\em A\&A} {\bf 2017}, {\em 600},~A124,
  \href{http://xxx.lanl.gov/abs/1702.07730}{{\normalfont
  [arXiv:astro-ph.HE/1702.07730]}}.
\newblock {\url{https://doi.org/10.1051/0004-6361/201630092}}.

\bibitem[Steinberg and Stone(2022)]{Steinberg20221}
Steinberg, E.; Stone, N.C.
\newblock The Origins of Peak Light in Tidal Disruption Events,  2022.
\newblock {\url{https://doi.org/10.48550/ARXIV.2206.10641}}.

\bibitem[{Stone} \em{et~al.}(2013){Stone}, {Sari}, and
  {Loeb}]{2013MNRAS.435.1809S}
{Stone}, N.; {Sari}, R.; {Loeb}, A.
\newblock {Consequences of strong compression in tidal disruption events}.
\newblock {\em MNRAS} {\bf 2013}, {\em 435},~1809--1824,
  \href{http://xxx.lanl.gov/abs/1210.3374}{{\normalfont
  [arXiv:astro-ph.HE/1210.3374]}}.
\newblock {\url{https://doi.org/10.1093/mnras/stt1270}}.

\bibitem[{Golightly} \em{et~al.}(2019){Golightly}, {Nixon}, and
  {Coughlin}]{2019ApJ...882L..26G}
{Golightly}, E.C.A.; {Nixon}, C.J.; {Coughlin}, E.R.
\newblock {On the Diversity of Fallback Rates from Tidal Disruption Events with
  Accurate Stellar Structure}.
\newblock {\em ApJ} {\bf 2019}, {\em 882},~L26,
  \href{http://xxx.lanl.gov/abs/1907.05895}{{\normalfont
  [arXiv:astro-ph.HE/1907.05895]}}.
\newblock {\url{https://doi.org/10.3847/2041-8213/ab380d}}.

\bibitem[{Komossa}(2015)]{2015JHEAp...7..148K}
{Komossa}, S.
\newblock {Tidal disruption of stars by supermassive black holes: Status of
  observations}.
\newblock {\em Journal of High Energy Astrophysics} {\bf 2015}, {\em
  7},~148--157,  \href{http://xxx.lanl.gov/abs/1505.01093}{{\normalfont
  [arXiv:astro-ph.HE/1505.01093]}}.
\newblock {\url{https://doi.org/10.1016/j.jheap.2015.04.006}}.

\bibitem[{Gezari}(2021)]{2021ARA&A..59...21G}
{Gezari}, S.
\newblock {Tidal Disruption Events}.
\newblock {\em ARA\&A} {\bf 2021}, {\em 59},
  \href{http://xxx.lanl.gov/abs/2104.14580}{{\normalfont
  [arXiv:astro-ph.HE/2104.14580]}}.
\newblock {\url{https://doi.org/10.1146/annurev-astro-111720-030029}}.

\bibitem[{Alexander} \em{et~al.}(2020){Alexander}, {van Velzen}, {Horesh}, and
  {Zauderer}]{2020SSRv..216...81A}
{Alexander}, K.D.; {van Velzen}, S.; {Horesh}, A.; {Zauderer}, B.A.
\newblock {Radio Properties of Tidal Disruption Events}.
\newblock {\em Space Sci. Rev.} {\bf 2020}, {\em 216},~81,
  \href{http://xxx.lanl.gov/abs/2006.01159}{{\normalfont
  [arXiv:astro-ph.HE/2006.01159]}}.
\newblock {\url{https://doi.org/10.1007/s11214-020-00702-w}}.

\bibitem[{Dai} \em{et~al.}(2018){Dai}, {McKinney}, {Roth}, {Ramirez-Ruiz}, and
  {Miller}]{2018ApJ...859L..20D}
{Dai}, L.; {McKinney}, J.C.; {Roth}, N.; {Ramirez-Ruiz}, E.; {Miller}, M.C.
\newblock {A Unified Model for Tidal Disruption Events}.
\newblock {\em ApJ} {\bf 2018}, {\em 859},~L20,
  \href{http://xxx.lanl.gov/abs/1803.03265}{{\normalfont
  [arXiv:astro-ph.HE/1803.03265]}}.
\newblock {\url{https://doi.org/10.3847/2041-8213/aab429}}.

\bibitem[{Curd} \em{et~al.}(2022){Curd}, {Emami}, {Anantua}, {Palumbo},
  {Doeleman}, and {Narayan}]{2022arXiv220606358C}
{Curd}, B.; {Emami}, R.; {Anantua}, R.; {Palumbo}, D.; {Doeleman}, S.;
  {Narayan}, R.
\newblock {Jets from SANE Super-Eddington Accretion Disks: Morphology, Spectra,
  and Their Potential as Targets for ngEHT}.
\newblock {\em arXiv e-prints} {\bf 2022}, p. arXiv:2206.06358,
  \href{http://xxx.lanl.gov/abs/2206.06358}{{\normalfont
  [arXiv:astro-ph.HE/2206.06358]}}.

\bibitem[{Gammie} \em{et~al.}(2003){Gammie}, {McKinney}, and
  {T{\'o}th}]{2003ApJ...589..444G}
{Gammie}, C.F.; {McKinney}, J.C.; {T{\'o}th}, G.
\newblock {HARM: A Numerical Scheme for General Relativistic
  Magnetohydrodynamics}.
\newblock {\em ApJ} {\bf 2003}, {\em 589},~444--457,
  \href{http://xxx.lanl.gov/abs/astro-ph/0301509}{{\normalfont
  [arXiv:astro-ph/astro-ph/0301509]}}.
\newblock {\url{https://doi.org/10.1086/374594}}.

\bibitem[{Tchekhovskoy} \em{et~al.}(2014){Tchekhovskoy}, {Metzger}, {Giannios},
  and {Kelley}]{2014MNRAS.437.2744T}
{Tchekhovskoy}, A.; {Metzger}, B.D.; {Giannios}, D.; {Kelley}, L.Z.
\newblock {Swift J1644+57 gone MAD: the case for dynamically important magnetic
  flux threading the black hole in a jetted tidal disruption event}.
\newblock {\em MNRAS} {\bf 2014}, {\em 437},~2744--2760,
  \href{http://xxx.lanl.gov/abs/1301.1982}{{\normalfont
  [arXiv:astro-ph.HE/1301.1982]}}.
\newblock {\url{https://doi.org/10.1093/mnras/stt2085}}.

\bibitem[{Curd} and {Narayan}(2019)]{2019MNRAS.483..565C}
{Curd}, B.; {Narayan}, R.
\newblock {GRRMHD simulations of tidal disruption event accretion discs around
  supermassive black holes: jet formation, spectra, and detectability}.
\newblock {\em MNRAS} {\bf 2019}, {\em 483},~565--592,
  \href{http://xxx.lanl.gov/abs/1811.06971}{{\normalfont
  [arXiv:astro-ph.HE/1811.06971]}}.
\newblock {\url{https://doi.org/10.1093/mnras/sty3134}}.

\bibitem[{Blandford} and {Znajek}(1977)]{1977MNRAS.179..433B}
{Blandford}, R.D.; {Znajek}, R.L.
\newblock {Electromagnetic extraction of energy from Kerr black holes.}
\newblock {\em MNRAS} {\bf 1977}, {\em 179},~433--456.
\newblock {\url{https://doi.org/10.1093/mnras/179.3.433}}.

\bibitem[{Coughlin} and {Begelman}(2020)]{2020MNRAS.499.3158C}
{Coughlin}, E.R.; {Begelman}, M.C.
\newblock {Structured, relativistic jets driven by radiation}.
\newblock {\em MNRAS} {\bf 2020}, {\em 499},~3158--3177,
  \href{http://xxx.lanl.gov/abs/2009.03898}{{\normalfont
  [arXiv:astro-ph.HE/2009.03898]}}.
\newblock {\url{https://doi.org/10.1093/mnras/staa3026}}.

\bibitem[{Thomsen} \em{et~al.}(2022){Thomsen}, {Kwan}, {Dai}, {Wu}, {Roth}, and
  {Ramirez-Ruiz}]{2022ApJ...937L..28T}
{Thomsen}, L.L.; {Kwan}, T.M.; {Dai}, L.; {Wu}, S.C.; {Roth}, N.;
  {Ramirez-Ruiz}, E.
\newblock {Dynamical Unification of Tidal Disruption Events}.
\newblock {\em ApJ} {\bf 2022}, {\em 937},~L28,
  \href{http://xxx.lanl.gov/abs/2206.02804}{{\normalfont
  [arXiv:astro-ph.HE/2206.02804]}}.
\newblock {\url{https://doi.org/10.3847/2041-8213/ac911f}}.

\bibitem[{Doeleman} \em{et~al.}(2019){Doeleman}, {Blackburn}, {Dexter},
  {Gomez}, {Johnson}, {Palumbo}, {Weintroub}, {Farah}, {Fish}, {Loinard},
  {Lonsdale}, {Narayanan}, {Patel}, {Pesce}, {Raymond}, {Tilanus}, {Wielgus},
  {Akiyama}, {Bower}, {Broderick}, {Deane}, {Fromm}, {Gammie}, {Gold},
  {Janssen}, {Kawashima}, {Krichbaum}, {Marrone}, {Matthews}, {Mizuno},
  {Rezzolla}, {Roelofs}, {Ros}, {Savolainen}, {Yuan}, {Zhao}, {Blackburn},
  {Doeleman}, {Dexter}, {Gomez}, {Johnson}, {Palumbo}, {Weintroub}, {Farah},
  {Fish}, {Loinard}, {Lonsdale}, {Narayanan}, {Patel}, {Pesce}, {Raymond},
  {Tilanus}, {Wielgus}, {Akiyama}, {Bower}, {Broderick}, {Deane}, {Fromm},
  {Gammie}, {Gold}, {Janssen}, {Kawashima}, {Krichbaum}, {Marrone}, {Matthews},
  {Mizuno}, {Rezzolla}, {Roelofs}, {Ros}, {Savolainen}, {Yuan}, and
  {Zhao}]{2019BAAS...51g.256D}
{Doeleman}, S.; {Blackburn}, L.; {Dexter}, J.; {Gomez}, J.L.; {Johnson}, M.D.;
  {Palumbo}, D.C.; {Weintroub}, J.; {Farah}, J.R.; {Fish}, V.; {Loinard}, L.;
  et~al.
\newblock {Studying Black Holes on Horizon Scales with VLBI Ground Arrays}.
\newblock In Proceedings of the Bulletin of the American Astronomical Society,
  2019, Vol.~51, p. 256,
  \href{http://xxx.lanl.gov/abs/1909.01411}{{\normalfont
  [arXiv:astro-ph.IM/1909.01411]}}.

\bibitem[{Ivezi{\'c}} \em{et~al.}(2019){Ivezi{\'c}}, {Kahn}, {Tyson}, {Abel},
  {Acosta}, {Allsman}, {Alonso}, {AlSayyad}, {Anderson}, {Andrew}, {Angel},
  {Angeli}, {Ansari}, {Antilogus}, {Araujo}, {Armstrong}, {Arndt}, {Astier},
  {Aubourg}, {Auza}, {Axelrod}, {Bard}, {Barr}, {Barrau}, {Bartlett}, {Bauer},
  {Bauman}, {Baumont}, {Bechtol}, {Bechtol}, {Becker}, {Becla}, {Beldica},
  {Bellavia}, {Bianco}, {Biswas}, {Blanc}, {Blazek}, {Blandford}, {Bloom},
  {Bogart}, {Bond}, {Booth}, {Borgland}, {Borne}, {Bosch}, {Boutigny},
  {Brackett}, {Bradshaw}, {Brandt}, {Brown}, {Bullock}, {Burchat}, {Burke},
  {Cagnoli}, {Calabrese}, {Callahan}, {Callen}, {Carlin}, {Carlson},
  {Chandrasekharan}, {Charles-Emerson}, {Chesley}, {Cheu}, {Chiang}, {Chiang},
  {Chirino}, {Chow}, {Ciardi}, {Claver}, {Cohen-Tanugi}, {Cockrum}, {Coles},
  {Connolly}, {Cook}, {Cooray}, {Covey}, {Cribbs}, {Cui}, {Cutri}, {Daly},
  {Daniel}, {Daruich}, {Daubard}, {Daues}, {Dawson}, {Delgado}, {Dellapenna},
  {de Peyster}, {de Val-Borro}, {Digel}, {Doherty}, {Dubois},
  {Dubois-Felsmann}, {Durech}, {Economou}, {Eifler}, {Eracleous}, {Emmons},
  {Fausti Neto}, {Ferguson}, {Figueroa}, {Fisher-Levine}, {Focke}, {Foss},
  {Frank}, {Freemon}, {Gangler}, {Gawiser}, {Geary}, {Gee}, {Geha}, {Gessner},
  {Gibson}, {Gilmore}, {Glanzman}, {Glick}, {Goldina}, {Goldstein}, {Goodenow},
  {Graham}, {Gressler}, {Gris}, {Guy}, {Guyonnet}, {Haller}, {Harris},
  {Hascall}, {Haupt}, {Hernandez}, {Herrmann}, {Hileman}, {Hoblitt}, {Hodgson},
  {Hogan}, {Howard}, {Huang}, {Huffer}, {Ingraham}, {Innes}, {Jacoby}, {Jain},
  {Jammes}, {Jee}, {Jenness}, {Jernigan}, {Jevremovi{\'c}}, {Johns}, {Johnson},
  {Johnson}, {Jones}, {Juramy-Gilles}, {Juri{\'c}}, {Kalirai}, {Kallivayalil},
  {Kalmbach}, {Kantor}, {Karst}, {Kasliwal}, {Kelly}, {Kessler}, {Kinnison},
  {Kirkby}, {Knox}, {Kotov}, {Krabbendam}, {Krughoff}, {Kub{\'a}nek},
  {Kuczewski}, {Kulkarni}, {Ku}, {Kurita}, {Lage}, {Lambert}, {Lange},
  {Langton}, {Le Guillou}, {Levine}, {Liang}, {Lim}, {Lintott}, {Long},
  {Lopez}, {Lotz}, {Lupton}, {Lust}, {MacArthur}, {Mahabal}, {Mandelbaum},
  {Markiewicz}, {Marsh}, {Marshall}, {Marshall}, {May}, {McKercher}, {McQueen},
  {Meyers}, {Migliore}, {Miller}, {Mills}, {Miraval}, {Moeyens}, {Moolekamp},
  {Monet}, {Moniez}, {Monkewitz}, {Montgomery}, {Morrison}, {Mueller},
  {Muller}, {Mu{\~n}oz Arancibia}, {Neill}, {Newbry}, {Nief}, {Nomerotski},
  {Nordby}, {O'Connor}, {Oliver}, {Olivier}, {Olsen}, {O'Mullane}, {Ortiz},
  {Osier}, {Owen}, {Pain}, {Palecek}, {Parejko}, {Parsons}, {Pease},
  {Peterson}, {Peterson}, {Petravick}, {Libby Petrick}, {Petry},
  {Pierfederici}, {Pietrowicz}, {Pike}, {Pinto}, {Plante}, {Plate}, {Plutchak},
  {Price}, {Prouza}, {Radeka}, {Rajagopal}, {Rasmussen}, {Regnault}, {Reil},
  {Reiss}, {Reuter}, {Ridgway}, {Riot}, {Ritz}, {Robinson}, {Roby}, {Roodman},
  {Rosing}, {Roucelle}, {Rumore}, {Russo}, {Saha}, {Sassolas}, {Schalk},
  {Schellart}, {Schindler}, {Schmidt}, {Schneider}, {Schneider}, {Schoening},
  {Schumacher}, {Schwamb}, {Sebag}, {Selvy}, {Sembroski}, {Seppala}, {Serio},
  {Serrano}, {Shaw}, {Shipsey}, {Sick}, {Silvestri}, {Slater}, {Smith},
  {Smith}, {Sobhani}, {Soldahl}, {Storrie-Lombardi}, {Stover}, {Strauss},
  {Street}, {Stubbs}, {Sullivan}, {Sweeney}, {Swinbank}, {Szalay}, {Takacs},
  {Tether}, {Thaler}, {Thayer}, {Thomas}, {Thornton}, {Thukral}, {Tice},
  {Trilling}, {Turri}, {Van Berg}, {Vanden Berk}, {Vetter}, {Virieux},
  {Vucina}, {Wahl}, {Walkowicz}, {Walsh}, {Walter}, {Wang}, {Wang}, {Warner},
  {Wiecha}, {Willman}, {Winters}, {Wittman}, {Wolff}, {Wood-Vasey}, {Wu},
  {Xin}, {Yoachim}, and {Zhan}]{2019ApJ...873..111I}
{Ivezi{\'c}}, {\v{Z}}.; {Kahn}, S.M.; {Tyson}, J.A.; {Abel}, B.; {Acosta}, E.;
  {Allsman}, R.; {Alonso}, D.; {AlSayyad}, Y.; {Anderson}, S.F.; {Andrew}, J.;
  et~al.
\newblock {LSST: From Science Drivers to Reference Design and Anticipated Data
  Products}.
\newblock {\em ApJ} {\bf 2019}, {\em 873},~111,
  \href{http://xxx.lanl.gov/abs/0805.2366}{{\normalfont
  [arXiv:astro-ph/0805.2366]}}.
\newblock {\url{https://doi.org/10.3847/1538-4357/ab042c}}.

\bibitem[{Bricman} and {Gomboc}(2020)]{2020ApJ...890...73B}
{Bricman}, K.; {Gomboc}, A.
\newblock {The Prospects of Observing Tidal Disruption Events with the Large
  Synoptic Survey Telescope}.
\newblock {\em ApJ} {\bf 2020}, {\em 890},~73,
  \href{http://xxx.lanl.gov/abs/1906.08235}{{\normalfont
  [arXiv:astro-ph.HE/1906.08235]}}.
\newblock {\url{https://doi.org/10.3847/1538-4357/ab6989}}.

\bibitem[{Stone} and {Metzger}(2016)]{2016MNRAS.455..859S}
{Stone}, N.C.; {Metzger}, B.D.
\newblock {Rates of stellar tidal disruption as probes of the supermassive
  black hole mass function}.
\newblock {\em MNRAS} {\bf 2016}, {\em 455},~859--883,
  \href{http://xxx.lanl.gov/abs/1410.7772}{{\normalfont
  [arXiv:astro-ph.HE/1410.7772]}}.
\newblock {\url{https://doi.org/10.1093/mnras/stv2281}}.

\bibitem[Sadowski \em{et~al.}(2016)Sadowski, Tejeda, Gafton, Rosswog, and
  Abarca]{Sadowski2016}
Sadowski, A.; Tejeda, E.; Gafton, E.; Rosswog, S.; Abarca, D.
\newblock {Magnetohydrodynamical simulations of a deep tidal disruption in
  general relativity}.
\newblock {\em Mon. Not. Roy. Astron. Soc.} {\bf 2016}, {\em 458},~4250--4268,
  \href{http://xxx.lanl.gov/abs/1512.04865}{{\normalfont
  [arXiv:astro-ph.HE/1512.04865]}}.
\newblock {\url{https://doi.org/10.1093/mnras/stw589}}.

\bibitem[Curd(2021)]{Curd2021}
Curd, B.
\newblock {Global simulations of tidal disruption event disc formation via
  stream injection in GRRMHD}.
\newblock {\em Mon. Not. Roy. Astron. Soc.} {\bf 2021}, {\em 507},~3207--3227,
  \href{http://xxx.lanl.gov/abs/2105.09904}{{\normalfont
  [arXiv:astro-ph.HE/2105.09904]}}.
\newblock {\url{https://doi.org/10.1093/mnras/stab2172}}.

\bibitem[{Novikov} and {Thorne}(1973)]{1973blho.conf..343N}
{Novikov}, I.D.; {Thorne}, K.S.
\newblock {Astrophysics of black holes.}
\newblock In Proceedings of the Black Holes (Les Astres Occlus),  1973, pp.
  343--450.

\bibitem[{Mo{\'s}cibrodzka} and {Gammie}(2018)]{Moscibrodzka2018}
{Mo{\'s}cibrodzka}, M.; {Gammie}, C.F.
\newblock {IPOLE - semi-analytic scheme for relativistic polarized radiative
  transport}.
\newblock {\em MNRAS} {\bf 2018}, {\em 475},~43--54,
  \href{http://xxx.lanl.gov/abs/1712.03057}{{\normalfont
  [arXiv:astro-ph.HE/1712.03057]}}.
\newblock {\url{https://doi.org/10.1093/mnras/stx3162}}.

\bibitem[{Yarza} \em{et~al.}(2020){Yarza}, {Wong}, {Ryan}, and
  {Gammie}]{Yarza2020}
{Yarza}, R.; {Wong}, G.N.; {Ryan}, B.R.; {Gammie}, C.F.
\newblock {Bremsstrahlung in GRMHD Models of Accreting Black Holes}.
\newblock {\em ApJ} {\bf 2020}, {\em 898},~50,
  \href{http://xxx.lanl.gov/abs/2006.01145}{{\normalfont
  [arXiv:astro-ph.HE/2006.01145]}}.
\newblock {\url{https://doi.org/10.3847/1538-4357/ab9808}}.

\bibitem[{Wong} \em{et~al.}(2022){Wong}, {Prather}, {Dhruv}, {Ryan},
  {Mo{\'s}cibrodzka}, {Chan}, {Joshi}, {Yarza}, {Ricarte}, {Shiokawa},
  {Dolence}, {Noble}, {McKinney}, and {Gammie}]{Wong2022}
{Wong}, G.N.; {Prather}, B.S.; {Dhruv}, V.; {Ryan}, B.R.; {Mo{\'s}cibrodzka},
  M.; {Chan}, C.k.; {Joshi}, A.V.; {Yarza}, R.; {Ricarte}, A.; {Shiokawa}, H.;
  et~al.
\newblock {PATOKA: Simulating Electromagnetic Observables of Black Hole
  Accretion}.
\newblock {\em ApJS} {\bf 2022}, {\em 259},~64,
  \href{http://xxx.lanl.gov/abs/2202.11721}{{\normalfont
  [arXiv:astro-ph.HE/2202.11721]}}.
\newblock {\url{https://doi.org/10.3847/1538-4365/ac582e}}.

\bibitem[{Ohmura} \em{et~al.}(2019){Ohmura}, {Machida}, {Nakamura}, {Kudoh},
  {Asahina}, and {Matsumoto}]{2019Galax...7...14O}
{Ohmura}, T.; {Machida}, M.; {Nakamura}, K.; {Kudoh}, Y.; {Asahina}, Y.;
  {Matsumoto}, R.
\newblock {Two-Temperature Magnetohydrodynamics Simulations of Propagation of
  Semi-Relativistic Jets}.
\newblock {\em Galaxies} {\bf 2019}, {\em 7},~14.
\newblock {\url{https://doi.org/10.3390/galaxies7010014}}.

\bibitem[{Ohmura} \em{et~al.}(2020){Ohmura}, {Machida}, {Nakamura}, {Kudoh},
  and {Matsumoto}]{2020MNRAS.493.5761O}
{Ohmura}, T.; {Machida}, M.; {Nakamura}, K.; {Kudoh}, Y.; {Matsumoto}, R.
\newblock {Two-temperature magnetohydrodynamic simulations for sub-relativistic
  active galactic nucleus jets: dependence on the fraction of the electron
  heating}.
\newblock {\em MNRAS} {\bf 2020}, {\em 493},~5761--5772,
  \href{http://xxx.lanl.gov/abs/2003.00795}{{\normalfont
  [arXiv:astro-ph.HE/2003.00795]}}.
\newblock {\url{https://doi.org/10.1093/mnras/staa632}}.

\bibitem[{Raymond} \em{et~al.}(2021){Raymond}, {Palumbo}, {Paine}, {Blackburn},
  {C{\'o}rdova Rosado}, {Doeleman}, {Farah}, {Johnson}, {Roelofs}, {Tilanus},
  and {Weintroub}]{Raymond2021}
{Raymond}, A.W.; {Palumbo}, D.; {Paine}, S.N.; {Blackburn}, L.; {C{\'o}rdova
  Rosado}, R.; {Doeleman}, S.S.; {Farah}, J.R.; {Johnson}, M.D.; {Roelofs}, F.;
  {Tilanus}, R.P.J.;  et~al.
\newblock {Evaluation of New Submillimeter VLBI Sites for the Event Horizon
  Telescope}.
\newblock {\em ApJS} {\bf 2021}, {\em 253},~5,
  \href{http://xxx.lanl.gov/abs/2102.05482}{{\normalfont
  [arXiv:astro-ph.IM/2102.05482]}}.
\newblock {\url{https://doi.org/10.3847/1538-3881/abc3c3}}.

\bibitem[Roelofs \em{et~al.}(2022, in prep.)Roelofs et~al.]{Roelofs2022}
Roelofs, F.;  et~al.,  2022, in prep.

\bibitem[{Chael} \em{et~al.}(2016){Chael}, {Johnson}, {Narayan}, {Doeleman},
  {Wardle}, and {Bouman}]{Chael2016}
{Chael}, A.A.; {Johnson}, M.D.; {Narayan}, R.; {Doeleman}, S.S.; {Wardle},
  J.F.C.; {Bouman}, K.L.
\newblock {High-resolution Linear Polarimetric Imaging for the Event Horizon
  Telescope}.
\newblock {\em ApJ} {\bf 2016}, {\em 829},~11,
  \href{http://xxx.lanl.gov/abs/1605.06156}{{\normalfont
  [arXiv:astro-ph.IM/1605.06156]}}.
\newblock {\url{https://doi.org/10.3847/0004-637X/829/1/11}}.

\bibitem[{Chael} \em{et~al.}(2018){Chael}, {Johnson}, {Bouman}, {Blackburn},
  {Akiyama}, and {Narayan}]{Chael2018}
{Chael}, A.A.; {Johnson}, M.D.; {Bouman}, K.L.; {Blackburn}, L.L.; {Akiyama},
  K.; {Narayan}, R.
\newblock {Interferometric Imaging Directly with Closure Phases and Closure
  Amplitudes}.
\newblock {\em ApJ} {\bf 2018}, {\em 857},~23,
  \href{http://xxx.lanl.gov/abs/1803.07088}{{\normalfont
  [arXiv:astro-ph.IM/1803.07088]}}.
\newblock {\url{https://doi.org/10.3847/1538-4357/aab6a8}}.

\bibitem[Doeleman \em{et~al.}(2022, in prep.)Doeleman et~al.]{Doeleman2022}
Doeleman, S.;  et~al.,  2022, in prep.

\bibitem[Gelaro \em{et~al.}(2017)Gelaro, McCarty, Suárez, Todling, Molod,
  Takacs, Randles, Darmenov, Bosilovich, Reichle, Wargan, Coy, Cullather,
  Draper, Akella, Buchard, Conaty, da~Silva, Gu, Kim, Koster, Lucchesi,
  Merkova, Nielsen, Partyka, Pawson, Putman, Rienecker, Schubert, Sienkiewicz,
  and Zhao]{Gelaro2017}
Gelaro, R.; McCarty, W.; Suárez, M.J.; Todling, R.; Molod, A.; Takacs, L.;
  Randles, C.A.; Darmenov, A.; Bosilovich, M.G.; Reichle, R.;  et~al.
\newblock {The Modern-Era Retrospective Analysis for Research and Applications,
  Version 2 (MERRA-2)}.
\newblock {\em Journal of Climate} {\bf 2017}, {\em 30},~5419--5454,
  \href{http://xxx.lanl.gov/abs/https://journals.ametsoc.org/jcli/article-pdf/30/14/5419/4676847/jcli-d-16-0758\_1.pdf}{{\normalfont
  [https://journals.ametsoc.org/jcli/article-pdf/30/14/5419/4676847/jcli-d-16-0758\_1.pdf]}}.
\newblock {\url{https://doi.org/10.1175/JCLI-D-16-0758.1}}.

\bibitem[Paine(2019)]{Paine2019}
Paine, S.
\newblock The am atmospheric model,  2019.
\newblock {\url{https://doi.org/10.5281/zenodo.3406496}}.

\bibitem[{Event Horizon Telescope Collaboration} \em{et~al.}(2019){Event
  Horizon Telescope Collaboration}, {Akiyama}, {Alberdi}, {Alef}, {Asada},
  {Azulay}, {Baczko}, {Ball}, {Balokovi{\'c}}, {Barrett}, {Bintley},
  {Blackburn}, {Boland}, {Bouman}, {Bower}, {Bremer}, {Brinkerink},
  {Brissenden}, {Britzen}, {Broderick}, {Broguiere}, {Bronzwaer}, {Byun},
  {Carlstrom}, {Chael}, {Chan}, {Chatterjee}, {Chatterjee}, {Chen}, {Chen},
  {Cho}, {Christian}, {Conway}, {Cordes}, {Crew}, {Cui}, {Davelaar}, {De
  Laurentis}, {Deane}, {Dempsey}, {Desvignes}, {Dexter}, {Doeleman}, {Eatough},
  {Falcke}, {Fish}, {Fomalont}, {Fraga-Encinas}, {Freeman}, {Friberg}, {Fromm},
  {G{\'o}mez}, {Galison}, {Gammie}, {Garc{\'\i}a}, {Gentaz}, {Georgiev},
  {Goddi}, {Gold}, {Gu}, {Gurwell}, {Hada}, {Hecht}, {Hesper}, {Ho}, {Ho},
  {Honma}, {Huang}, {Huang}, {Hughes}, {Ikeda}, {Inoue}, {Issaoun}, {James},
  {Jannuzi}, {Janssen}, {Jeter}, {Jiang}, {Johnson}, {Jorstad}, {Jung},
  {Karami}, {Karuppusamy}, {Kawashima}, {Keating}, {Kettenis}, {Kim}, {Kim},
  {Kim}, {Kino}, {Koay}, {Koch}, {Koyama}, {Kramer}, {Kramer}, {Krichbaum},
  {Kuo}, {Lauer}, {Lee}, {Li}, {Li}, {Lindqvist}, {Liu}, {Liuzzo}, {Lo},
  {Lobanov}, {Loinard}, {Lonsdale}, {Lu}, {MacDonald}, {Mao}, {Markoff},
  {Marrone}, {Marscher}, {Mart{\'\i}-Vidal}, {Matsushita}, {Matthews},
  {Medeiros}, {Menten}, {Mizuno}, {Mizuno}, {Moran}, {Moriyama},
  {Moscibrodzka}, {M{\"u}ller}, {Nagai}, {Nagar}, {Nakamura}, {Narayan},
  {Narayanan}, {Natarajan}, {Neri}, {Ni}, {Noutsos}, {Okino}, {Olivares},
  {Oyama}, {{\"O}zel}, {Palumbo}, {Patel}, {Pen}, {Pesce}, {Pi{\'e}tu},
  {Plambeck}, {PopStefanija}, {Porth}, {Prather}, {Preciado-L{\'o}pez},
  {Psaltis}, {Pu}, {Ramakrishnan}, {Rao}, {Rawlings}, {Raymond}, {Rezzolla},
  {Ripperda}, {Roelofs}, {Rogers}, {Ros}, {Rose}, {Roshanineshat}, {Rottmann},
  {Roy}, {Ruszczyk}, {Ryan}, {Rygl}, {S{\'a}nchez}, {S{\'a}nchez-Arguelles},
  {Sasada}, {Savolainen}, {Schloerb}, {Schuster}, {Shao}, {Shen}, {Small},
  {Sohn}, {SooHoo}, {Tazaki}, {Tiede}, {Tilanus}, {Titus}, {Toma}, {Torne},
  {Trent}, {Trippe}, {Tsuda}, {van Bemmel}, {van Langevelde}, {van Rossum},
  {Wagner}, {Wardle}, {Weintroub}, {Wex}, {Wharton}, {Wielgus}, {Wong}, {Wu},
  {Young}, {Young}, {Younsi}, {Yuan}, {Yuan}, {Zensus}, {Zhao}, {Zhao}, {Zhu},
  {Farah}, {Meyer-Zhao}, {Michalik}, {Nadolski}, {Nishioka}, {Pradel},
  {Primiani}, {Souccar}, {Vertatschitsch}, and {Yamaguchi}]{EHT2019M87IV}
{Event Horizon Telescope Collaboration}.; {Akiyama}, K.; {Alberdi}, A.; {Alef},
  W.; {Asada}, K.; {Azulay}, R.; {Baczko}, A.K.; {Ball}, D.; {Balokovi{\'c}},
  M.; {Barrett}, J.;  et~al.
\newblock {First M87 Event Horizon Telescope Results. IV. Imaging the Central
  Supermassive Black Hole}.
\newblock {\em ApJ} {\bf 2019}, {\em 875},~L4,
  \href{http://xxx.lanl.gov/abs/1906.11241}{{\normalfont
  [arXiv:astro-ph.GA/1906.11241]}}.
\newblock {\url{https://doi.org/10.3847/2041-8213/ab0e85}}.

\bibitem[{Jorstad} \em{et~al.}(2005){Jorstad}, {Marscher}, {Lister},
  {Stirling}, {Cawthorne}, {Gear}, {G{\'o}mez}, {Stevens}, {Smith}, {Forster},
  and {Robson}]{2005AJ....130.1418J}
{Jorstad}, S.G.; {Marscher}, A.P.; {Lister}, M.L.; {Stirling}, A.M.;
  {Cawthorne}, T.V.; {Gear}, W.K.; {G{\'o}mez}, J.L.; {Stevens}, J.A.; {Smith},
  P.S.; {Forster}, J.R.;  et~al.
\newblock {Polarimetric Observations of 15 Active Galactic Nuclei at High
  Frequencies: Jet Kinematics from Bimonthly Monitoring with the Very Long
  Baseline Array}.
\newblock {\em AJ} {\bf 2005}, {\em 130},~1418--1465,
  \href{http://xxx.lanl.gov/abs/astro-ph/0502501}{{\normalfont
  [arXiv:astro-ph/astro-ph/0502501]}}.
\newblock {\url{https://doi.org/10.1086/444593}}.

\bibitem[{Lister} \em{et~al.}(2013){Lister}, {Aller}, {Aller}, {Homan},
  {Kellermann}, {Kovalev}, {Pushkarev}, {Richards}, {Ros}, and
  {Savolainen}]{2013AJ....146..120L}
{Lister}, M.L.; {Aller}, M.F.; {Aller}, H.D.; {Homan}, D.C.; {Kellermann},
  K.I.; {Kovalev}, Y.Y.; {Pushkarev}, A.B.; {Richards}, J.L.; {Ros}, E.;
  {Savolainen}, T.
\newblock {MOJAVE. X. Parsec-scale Jet Orientation Variations and Superluminal
  Motion in Active Galactic Nuclei}.
\newblock {\em AJ} {\bf 2013}, {\em 146},~120,
  \href{http://xxx.lanl.gov/abs/1308.2713}{{\normalfont
  [arXiv:astro-ph.CO/1308.2713]}}.
\newblock {\url{https://doi.org/10.1088/0004-6256/146/5/120}}.

\bibitem[{Cohen} \em{et~al.}(2014){Cohen}, {Meier}, {Arshakian}, {Homan},
  {Hovatta}, {Kovalev}, {Lister}, {Pushkarev}, {Richards}, and
  {Savolainen}]{2014ApJ...787..151C}
{Cohen}, M.H.; {Meier}, D.L.; {Arshakian}, T.G.; {Homan}, D.C.; {Hovatta}, T.;
  {Kovalev}, Y.Y.; {Lister}, M.L.; {Pushkarev}, A.B.; {Richards}, J.L.;
  {Savolainen}, T.
\newblock {Studies of the Jet in Bl Lacertae. I. Recollimation Shock and Moving
  Emission Features}.
\newblock {\em ApJ} {\bf 2014}, {\em 787},~151,
  \href{http://xxx.lanl.gov/abs/1404.0976}{{\normalfont
  [arXiv:astro-ph.HE/1404.0976]}}.
\newblock {\url{https://doi.org/10.1088/0004-637X/787/2/151}}.

\bibitem[{Kohler} \em{et~al.}(2012){Kohler}, {Begelman}, and
  {Beckwith}]{2012MNRAS.422.2282K}
{Kohler}, S.; {Begelman}, M.C.; {Beckwith}, K.
\newblock {Recollimation boundary layers in relativistic jets}.
\newblock {\em MNRAS} {\bf 2012}, {\em 422},~2282--2290,
  \href{http://xxx.lanl.gov/abs/1112.4843}{{\normalfont
  [arXiv:astro-ph.HE/1112.4843]}}.
\newblock {\url{https://doi.org/10.1111/j.1365-2966.2012.20776.x}}.

\bibitem[{Lazzati} \em{et~al.}(2012){Lazzati}, {Morsony}, {Blackwell}, and
  {Begelman}]{2012ApJ...750...68L}
{Lazzati}, D.; {Morsony}, B.J.; {Blackwell}, C.H.; {Begelman}, M.C.
\newblock {Unifying the Zoo of Jet-driven Stellar Explosions}.
\newblock {\em ApJ} {\bf 2012}, {\em 750},~68,
  \href{http://xxx.lanl.gov/abs/1111.0970}{{\normalfont
  [arXiv:astro-ph.HE/1111.0970]}}.
\newblock {\url{https://doi.org/10.1088/0004-637X/750/1/68}}.

\bibitem[{Mizuno} \em{et~al.}(2015){Mizuno}, {G{\'o}mez}, {Nishikawa}, {Meli},
  {Hardee}, and {Rezzolla}]{2015ApJ...809...38M}
{Mizuno}, Y.; {G{\'o}mez}, J.L.; {Nishikawa}, K.I.; {Meli}, A.; {Hardee}, P.E.;
  {Rezzolla}, L.
\newblock {Recollimation Shocks in Magnetized Relativistic Jets}.
\newblock {\em ApJ} {\bf 2015}, {\em 809},~38,
  \href{http://xxx.lanl.gov/abs/1505.00933}{{\normalfont
  [arXiv:astro-ph.HE/1505.00933]}}.
\newblock {\url{https://doi.org/10.1088/0004-637X/809/1/38}}.

\bibitem[{Hervet} \em{et~al.}(2017){Hervet}, {Meliani}, {Zech}, {Boisson},
  {Cayatte}, {Sauty}, and {Sol}]{2017A&A...606A.103H}
{Hervet}, O.; {Meliani}, Z.; {Zech}, A.; {Boisson}, C.; {Cayatte}, V.; {Sauty},
  C.; {Sol}, H.
\newblock {Shocks in relativistic transverse stratified jets. A new paradigm
  for radio-loud AGN}.
\newblock {\em A\&A} {\bf 2017}, {\em 606},~A103,
  \href{http://xxx.lanl.gov/abs/1705.10556}{{\normalfont
  [arXiv:astro-ph.HE/1705.10556]}}.
\newblock {\url{https://doi.org/10.1051/0004-6361/201730745}}.

\bibitem[{G{\'o}mez} \em{et~al.}(2016){G{\'o}mez}, {Lobanov}, {Bruni},
  {Kovalev}, {Marscher}, {Jorstad}, {Mizuno}, {Bach}, {Sokolovsky}, {Anderson},
  {Galindo}, {Kardashev}, and {Lisakov}]{2016ApJ...817...96G}
{G{\'o}mez}, J.L.; {Lobanov}, A.P.; {Bruni}, G.; {Kovalev}, Y.Y.; {Marscher},
  A.P.; {Jorstad}, S.G.; {Mizuno}, Y.; {Bach}, U.; {Sokolovsky}, K.V.;
  {Anderson}, J.M.;  et~al.
\newblock {Probing the Innermost Regions of AGN Jets and Their Magnetic Fields
  with RadioAstron. I. Imaging BL Lacertae at 21 Microarcsecond Resolution}.
\newblock {\em ApJ} {\bf 2016}, {\em 817},~96,
  \href{http://xxx.lanl.gov/abs/1512.04690}{{\normalfont
  [arXiv:astro-ph.HE/1512.04690]}}.
\newblock {\url{https://doi.org/10.3847/0004-637X/817/2/96}}.

\bibitem[{Zauderer} \em{et~al.}(2013){Zauderer}, {Berger}, {Margutti},
  {Pooley}, {Sari}, {Soderberg}, {Brunthaler}, and
  {Bietenholz}]{2013ApJ...767..152Z}
{Zauderer}, B.A.; {Berger}, E.; {Margutti}, R.; {Pooley}, G.G.; {Sari}, R.;
  {Soderberg}, A.M.; {Brunthaler}, A.; {Bietenholz}, M.F.
\newblock {Radio Monitoring of the Tidal Disruption Event Swift
  J164449.3+573451. II. The Relativistic Jet Shuts Off and a Transition to
  Forward Shock X-Ray/Radio Emission}.
\newblock {\em ApJ} {\bf 2013}, {\em 767},~152,
  \href{http://xxx.lanl.gov/abs/1212.1173}{{\normalfont
  [arXiv:astro-ph.HE/1212.1173]}}.
\newblock {\url{https://doi.org/10.1088/0004-637X/767/2/152}}.

\bibitem[{Pasham} \em{et~al.}(2015){Pasham}, {Cenko}, {Levan}, {Bower},
  {Horesh}, {Brown}, {Dolan}, {Wiersema}, {Filippenko}, {Fruchter}, {Greiner},
  {O'Brien}, {Page}, {Rau}, and {Tanvir}]{2015ApJ...805...68P}
{Pasham}, D.R.; {Cenko}, S.B.; {Levan}, A.J.; {Bower}, G.C.; {Horesh}, A.;
  {Brown}, G.C.; {Dolan}, S.; {Wiersema}, K.; {Filippenko}, A.V.; {Fruchter},
  A.S.;  et~al.
\newblock {A Multiwavelength Study of the Relativistic Tidal Disruption
  Candidate Swift J2058.4+0516 at Late Times}.
\newblock {\em ApJ} {\bf 2015}, {\em 805},~68,
  \href{http://xxx.lanl.gov/abs/1502.01345}{{\normalfont
  [arXiv:astro-ph.HE/1502.01345]}}.
\newblock {\url{https://doi.org/10.1088/0004-637X/805/1/68}}.

\bibitem[{Curd} and {Narayan}(2022)]{2022arXiv220912081C}
{Curd}, B.; {Narayan}, R.
\newblock {GRRMHD Simulations of MAD Accretion Disks Declining from
  Super-Eddington to Sub-Eddington Accretion Rates}.
\newblock {\em arXiv e-prints} {\bf 2022}, p. arXiv:2209.12081,
  \href{http://xxx.lanl.gov/abs/2209.12081}{{\normalfont
  [arXiv:astro-ph.HE/2209.12081]}}.

\bibitem[{Liska} \em{et~al.}(2022){Liska}, {Musoke}, {Tchekhovskoy}, {Porth},
  and {Beloborodov}]{2022ApJ...935L...1L}
{Liska}, M.T.P.; {Musoke}, G.; {Tchekhovskoy}, A.; {Porth}, O.; {Beloborodov},
  A.M.
\newblock {Formation of Magnetically Truncated Accretion Disks in 3D
  Radiation-transport Two-temperature GRMHD Simulations}.
\newblock {\em ApJ} {\bf 2022}, {\em 935},~L1,
  \href{http://xxx.lanl.gov/abs/2201.03526}{{\normalfont
  [arXiv:astro-ph.HE/2201.03526]}}.
\newblock {\url{https://doi.org/10.3847/2041-8213/ac84db}}.

\bibitem[{Cendes} \em{et~al.}(2022){Cendes}, {Berger}, {Alexander}, {Gomez},
  {Hajela}, {Chornock}, {Laskar}, {Margutti}, {Metzger}, {Bietenholz},
  {Brethauer}, and {Wieringa}]{2022ApJ...938...28C}
{Cendes}, Y.; {Berger}, E.; {Alexander}, K.D.; {Gomez}, S.; {Hajela}, A.;
  {Chornock}, R.; {Laskar}, T.; {Margutti}, R.; {Metzger}, B.; {Bietenholz},
  M.F.;  et~al.
\newblock {A Mildly Relativistic Outflow Launched Two Years after Disruption in
  Tidal Disruption Event AT2018hyz}.
\newblock {\em ApJ} {\bf 2022}, {\em 938},~28,
  \href{http://xxx.lanl.gov/abs/2206.14297}{{\normalfont
  [arXiv:astro-ph.HE/2206.14297]}}.
\newblock {\url{https://doi.org/10.3847/1538-4357/ac88d0}}.

\bibitem[Virtanen \em{et~al.}(2020)Virtanen, Gommers, Oliphant, Haberland,
  Reddy, Cournapeau, Burovski, Peterson, Weckesser, Bright, {van der Walt},
  Brett, Wilson, Millman, Mayorov, Nelson, Jones, Kern, Larson, Carey, Polat,
  Feng, Moore, {VanderPlas}, Laxalde, Perktold, Cimrman, Henriksen, Quintero,
  Harris, Archibald, Ribeiro, Pedregosa, {van Mulbregt}, and {SciPy 1.0
  Contributors}]{2020SciPy-NMeth}
Virtanen, P.; Gommers, R.; Oliphant, T.E.; Haberland, M.; Reddy, T.;
  Cournapeau, D.; Burovski, E.; Peterson, P.; Weckesser, W.; Bright, J.;
  et~al.
\newblock {{SciPy} 1.0: Fundamental Algorithms for Scientific Computing in
  Python}.
\newblock {\em Nature Methods} {\bf 2020}, {\em 17},~261--272.
\newblock {\url{https://doi.org/10.1038/s41592-019-0686-2}}.

\end{thebibliography}

\end{adjustwidth}
\end{document}